\documentclass[aps,
		prd,
		reprint,
		twocolumn,
		superscriptaddress,
		shortbibliography,
		nofootinbib,
		floatfix,
		notitlepage
		]{revtex4-1}
\pdfoutput=1

\usepackage{amsmath,amssymb}
\usepackage{graphicx,accents}
\usepackage{graphicx,epstopdf}
\usepackage[usenames,dvipsnames]{xcolor}
\usepackage{slashed}
\usepackage[normalem]{ulem}
\usepackage{bm}
\usepackage{bbm}
\usepackage{bbold}

\usepackage[printwatermark]{xwatermark}

\usepackage{hyperref}
\hypersetup{colorlinks=true}

\usepackage{tabularx}

\newcommand{\be}{\begin{equation}}
\newcommand{\ee}{\end{equation}}
\newcommand{\ba}{\begin{array}}
\newcommand{\ea}{\end{array}}
\newcommand{\bea}{\begin{eqnarray}}
\newcommand{\eea}{\end{eqnarray}}
\newcommand{\bd}{\begin{displaymath}}
\newcommand{\ed}{\end{displaymath}}

\newcommand{\trm}[1]{\textrm{#1}}

\newcommand{\vphi}{\varphi}

\newcommand{\ud}{\mathrm{d}}
\newcommand{\LCm}{{\scriptscriptstyle -}} 
\newcommand{\LCp}{{\scriptscriptstyle +}}
\newcommand{\LCpm}{{\scriptscriptstyle \pm}}

\newcommand{\maf}[1]{\mathfrak{#1}}
\newcommand{\LCperp}{{\scriptscriptstyle \perp}}
\newcommand{\bi}{\begin{itemize}}
\newcommand{\ei}{\end{itemize}}

\makeatletter
\DeclareRobustCommand{\cev}[1]{%
  \mathpalette\do@cev{#1}%
}
\newcommand{\do@cev}[2]{%
  \fix@cev{#1}{+}%
  \reflectbox{$\m@th#1\vec{\reflectbox{$\fix@cev{#1}{-}\m@th#1#2\fix@cev{#1}{+}$}}$}%
  \fix@cev{#1}{-}%
}
\newcommand{\fix@cev}[2]{%
  \ifx#1\displaystyle
    \mkern#23mu
  \else
    \ifx#1\textstyle
      \mkern#23mu
    \else
      \ifx#1\scriptstyle
        \mkern#22mu
      \else
        \mkern#22mu
      \fi
    \fi
  \fi
}

\newcommand*\xbar[1]{%
  \hbox{%
    \vbox{%
      \hrule height 0.5pt 
      \kern0.2ex
      \hbox{%
        \kern-0.15em
        \ensuremath{#1}%
        \kern-0.15em
      }%
    }%
  }%
}

\newcommand{\ST}[1]{#1}
\newcommand{\sscript}{\scriptscriptstyle}

\newcommand{\vsigma}{\varsigma}

\newcommand{\vtheta}{\vartheta}
\graphicspath{ {./figures/} }

\begin{document}
\title{Fully polarized nonlinear Breit-Wheeler pair production in pulsed plane waves}
\author{Suo Tang}
\email{tangsuo@ouc.edu.cn}
\affiliation{College of Physics and Optoelectronic Engineering, Ocean University of China, Qingdao, Shandong, 266100, China}



\begin{abstract}
Fully polarized nonlinear Breit-Wheeler pair production from a beam of polarized seed photons is investigated in pulsed plane-wave backgrounds.
The particle (electron and positron) spin and photon polarization are comprehensively described with the theory of the density matrix.
The compact expressions for the energy spectrum and spin polarization of the produced particles, depending on the initial polarization of the seed photon beam, are derived and discussed in both linearly and circularly polarized laser backgrounds.
The numerical results suggest an appreciable improvement of the positron yield by orthogonalizing the photon polarization to the laser polarization,
and the generation of a highly polarized positron beam from a beam of circularly polarized seed photons.
The locally monochromatic approximation and the locally constant field approximation are derived and benchmarked with the full quantum electrodynamic calculations.
\end{abstract}

\maketitle
%

\section{Introduction}
The decay of a high-energy photon into a pair of electron and positron when colliding with intense laser pulses,
is often referred to as the nonlinear Breit-Wheeler (NBW) pair production~\cite{breit34,Reiss1962,nikishov64},
and has been measured in the perturbative regime via the landmark E144 experiment more than two decades ago~\cite{burke97,bamber99}.
The measurement of the process in the transition regime from the perturbative to
the non-perturbative regime is one of the main goals for the modern-day laser-particle experiments~\cite{Abramowicz:2021zja,2021LUXEXFEL,macleod2021theory,slacref1,naranjo2021pair,E320_2021}. 
For the next generation of multi-PW laser facilities~\cite{Yoon:19,bromage2019,danson2019}, the NBW pair production will be explored in the thoroughly non-perturbative regime.

The properties of the produced electron and positron, such as the energy spectrum and transverse momentum distribution, have been theoretically investigated in various types of laser fields~\cite{nikishov64,TangPRA2021,PRA2012_052104,PRL2012_240406,PRA2013_062110,PRD013010,PRD2016_053011,PRA2013_052125,Titov:2015tdz,TangPRD096019,EPJD2020Titov,PRA2014052108,JANSEN201771,Titov:2018bgy,ilderton2019coherent,Ilderton:2019vot,king2020uniform};
monochromatic fields, constant (crossed) fields, as well as finite laser pulses.
When operating in a many-cycle laser pulse,
the energy spectrum of the produced pair could present discernible harmonic structure~\cite{nikishov64,TangPRA2021} if the centre-of-mass energy in the collision exceeds the threshold of $2mc^2$ with only a low number of laser photons,
where $m$ is electron (positron) rest mass and $c$ is the speed of light,
while colliding with short laser pulses could smooth the harmonic edges and lead to an asymmetric transverse momentum distribution even with circularly polarized laser pulses~\cite{PRA2013_052125,Titov:2015tdz,EPJD2020Titov,TangPRD096019}.
Pronounced interference phenomena could be manifested in the pair's spectrum and transverse momentum distribution when colliding with multi-pulse fields~\cite{PRA2014052108,JANSEN201771,Titov:2018bgy,ilderton2019coherent,Ilderton:2019vot}.

The spin- and polarization-dependence of the NBW process have also been broadly studied~\cite{ivanov2005complete,2010PRA062110,2011MULLER2011201,katkov2012production,PRD076017,Titov2020,SeiptPRA052805,chen2022electron} by, respectively, specifying a particular basis for the particle (electron and positron) spin and photon polarization.
It has been proved that the seed photon with different polarization could stimulate the pair production with different rate~\cite{breit34,PRD076017,Titov2020},
and the produced particles distribute on different spin bases with different probability~\cite{ivanov2005complete,SeiptPRA052805}.
However, the pair production rate from a particularly polarized photon \emph{cannot} thoroughly uncover the dependence of the process on the photon polarization,
and the probability for the produced particle to stay at one particular spin state \emph{cannot} reflect the full information of the particle spin.
A complete description of the photon polarization and fermion spin can be provided by the theory of the density matrix~\cite{blum2012density,balashov2013polarization}.
The density matrix has been well implemented in the nonlinear Compton scattering process to describe the spin polarization of the scattered electron~\cite{seipt18}.
The comprehensive description of the photon polarization effect and the particle spin distribution in the NBW process is still lacking, and is the main topic of this paper.
An alternative description of the particle polarization in quantum electrodynamic (QED) processes is the theory of Mueller matrix~\cite{Torgrimsson_2021}.

The study of the photon polarization effect on the \mbox{NBW} pair production and the spin polarization property of the produced particles is partly motivated by the upcoming high-energy laser-particle experiment, \emph{i.e.}, \mbox{LUXE} at \mbox{DESY}~\cite{Abramowicz:2021zja,2021LUXEXFEL,macleod2021theory} and \mbox{E320} at \mbox{SLAC}~\cite{slacref1,naranjo2021pair,E320_2021}
in which photons with energy $O(10~\trm{GeV})$ are planned to be generated to collide with laser pulses with the intermediate intensity $\xi\sim O(1)$, where $\xi$ is the classical nonlinearity parameter.
Compact expressions of the positron (electron) spectrum and spin polarization depending on the seed photon polarization are derived starting from the theory of scattering matrix and polarization density matrix.
The locally monochromatic approximation, used in simulations to capture the rich interference effects of relevant QED processes in the intermediate intensity regime~\cite{LMA063110,Blackburn_2021,blackburn2021higher}, is given,
and the locally constant field approximation, which has been broadly used in laser-plasma simulations~\cite{2011PRL035001,2011PRSTAB054401,TangPRA2014,2015PRE023305,TangPRA2019,WAN2020135120,Seipt_2021}, is also provided.

The paper is organised as follows.
In Sec.~\ref{Sec_2}, a general theoretical model is first presented for the NBW process from a beam of polarized seed photons to produce pairs of spin-polarized electron and positron, and then the theoretical model is applied explicitly in planed wave backgrounds.
In Sec.~\ref{Sec_3}, the energy spectrum and spin polarization of the produced positron are derived in planed wave backgrounds, and analyzed with the local approaches. 
The numerical implementations of the theoretical model and the benchmarking of the local approaches are preformed in Sec.~\ref{Sec_4} for differently polarized laser pulses.
At the end, the main points of the paper are concluded in Sec.~\ref{Sec_5}.
In following discussions, the natural units $\hbar=c=1$ is used, and the fine structure constant is $\alpha=e^2\approx1/137$.

\section{Theoretical model}~\label{Sec_2}
The NBW pair production is stimulated by a beam of seed $\gamma$-photons in the initial state:
\begin{align}
|\gamma;\ell,\varepsilon\rangle=|\gamma;\ell\rangle\otimes|\gamma;\varepsilon\rangle
\end{align}
which is characterised by a definite momentum (normalized) state $|\gamma;\ell\rangle$ and a polarization state $|\gamma;\varepsilon\rangle$.
In principle, the seed photon can stay either in a completely polarized pure state, or in a mixed state with partial polarization. To fully describe the polarization state of the photon beam, one can use the polarization density matrix~\cite{blum2012density,balashov2013polarization},
\begin{align}
|\gamma;\varepsilon\rangle\langle\gamma;\varepsilon|=\sum_{\lambda,\lambda'= \pm}\rho_{\gamma,\sscript\lambda\lambda'}|\varepsilon_{\lambda}\rangle\langle\varepsilon_{\lambda'}|
\end{align}
where $|\varepsilon_{\lambda}\rangle$ is an arbitrary set of polarization bases. Different polarization bases can only change the expression of $\rho_{\gamma}$, but not any relevant physical quantities.

The final density matrix of the NBW process can be given in form as
\begin{align}
|f\rangle\langle f| = \hat{S}|\gamma;\ell,\varepsilon\rangle\langle\gamma;\ell,\varepsilon|\hat{S}^{\dagger}
\end{align}
where the final state $|f\rangle=\hat{S}|\gamma;\ell,\varepsilon\rangle$ is comprised of a pair of electron and positron, and $\hat{S}$ is the corresponding scattering operator.
The spin polarization of the produced particles is manifested by tracing out their momentum states, as
\begin{widetext}
\begin{equation}
\rho_{\sscript f}=\int\frac{V\ud^{3}p}{(2\pi)^3}\frac{V\ud^{3}q}{(2\pi)^3} \begin{bmatrix}
\langle e^{\LCm};p,+| \\
\langle e^{\LCm};p,-|
\end{bmatrix}\otimes
\begin{bmatrix}
\langle e^{\LCp};q,+| \\
\langle e^{\LCp};q,-|
\end{bmatrix}|f\rangle\langle f|\begin{bmatrix}
|e^{\LCm};p,+\rangle & |e^{\LCm};p,-\rangle
\end{bmatrix}\otimes
\begin{bmatrix}
|e^{\LCp};q,+\rangle & |e^{\LCp};q,-\rangle
\end{bmatrix}\,,
\end{equation}
where $V$ is the volume factor, $|e^{\LCm};p,\sigma\rangle$ \mbox{($|e^{\LCp};q,\varsigma \rangle$)} denotes the eigenstate of the outgoing electron $e^{\LCm}$ (positron $e^{\LCp}$) with momentum $p^{\nu}$ ($q^{\nu}$) and spin $\sigma/2$ ($\varsigma/2$) along the quantum axis.
In principle, the spin quantum axis could also be selected arbitrarily.
$\rho_{\sscript f}$ is the polarization matrix involving the full information of the pair's spin states. With the direct calculation of $\rho_{\sscript f}$, one can uncover not only the spin polarization of each particle, but also the spin correlation between the paired particles.

In this work, we are interested in the spin properties of each produced particle, but not the spin correlation between the paired particles. 
This consideration is for the generation of a polarized particle source, and partially because of the difficulty in the coherent measurement for the spin state of the paired particles~\cite{salgado2021single}. 
To manifest the polarization properties of the produced positron $\rho^{\LCp}$ (electron $\rho^{\LCm}$), one can trace the matrix $\rho_{\sscript f}$ over the spin states of its paired electron (positron), as
\begin{subequations}
\begin{align}
\rho^{\LCp}=&\frac{1}{\trm{P}}\sum_{\lambda,\lambda'=\pm}\sum_{\sigma=\pm}\rho_{\gamma,\sscript\lambda\lambda'}\int\frac{V\ud^{3}p}{(2\pi)^3}\frac{V\ud^{3}q}{(2\pi)^3} \begin{bmatrix}
\langle p,\sigma;q,+| \\
\langle p,\sigma;q,-|
\end{bmatrix}\hat{S}\left|\ell,\varepsilon_{\lambda}\rangle\langle \ell,\varepsilon_{\lambda'}\right|\hat{S}^{\dagger}
\begin{bmatrix}
|p,\sigma;q,+\rangle & |p,\sigma;q,-\rangle
\end{bmatrix}\,,\\
\rho^{\LCm}=&\frac{1}{\trm{P}}\sum_{\lambda,\lambda'=\pm}\sum_{\varsigma=\pm}\rho_{\gamma,\sscript\lambda\lambda'}\int\frac{V\ud^{3}p}{(2\pi)^3}\frac{V\ud^{3}q}{(2\pi)^3}  \begin{bmatrix}
\langle p,+;q,~\varsigma| \\
\langle p,-;q,~\varsigma|
\end{bmatrix}\hat{S}\left|\ell,\varepsilon_{\lambda}\rangle\langle \ell,\varepsilon_{\lambda'}\right|\hat{S}^{\dagger}
\begin{bmatrix}
|p,+;q,\varsigma\rangle & |p,-;q,\varsigma\rangle
\end{bmatrix}\,,
\end{align}
\label{Eq_pair_spin_density}
\end{subequations}
\end{widetext}
where \mbox{$| p,\sigma;q,\varsigma\rangle=|e^{-};p,\sigma\rangle\otimes|e^{\LCp};q,\varsigma\rangle$} is the abbreviation of the composite state of the produced pair of electron and positron.
The spin polarization of the produced particle is convenient to be described with the spin density matrix~\cite{blum2012density,balashov2013polarization,seipt18},
given that the produced particle can stay not only in a completely polarized pure state,
but also in a partially polarized mixed state.
Here, one should realize that the direct product $\rho^{\LCp}\otimes \rho^{\LCm}$ may not reproduce the full polarization matrix $\rho_{\sscript f}$ as the spin correlation is eliminated by tracing out the spin states of one of the paired particles.
The spin density matrix $\rho^{\LCp}$ ($\rho^{\LCm}$) has been normalized with the total probability $\trm{P}$,
\begin{align}
\trm{P}&=\sum_{\lambda,\lambda'=\pm}\sum_{\sigma,\varsigma=\pm}\rho_{\gamma,\sscript\lambda\lambda'} \int\frac{V\ud^{3}p}{(2\pi)^3}\frac{V\ud^{3}q}{(2\pi)^3} \nonumber\\
&~~~\langle p,\sigma;q,\varsigma|\hat{S}|\ell,\varepsilon_{\lambda}\rangle\langle \ell,\varepsilon_{\lambda'}|\hat{S}^{\dagger}| p,\sigma;q,\varsigma\rangle\,,
\end{align}
which is acquired by tracing out all the spin states and uncovers the dependence of the NBW process on the polarization of the seed photon.

After calculating the scattering matrix element \mbox{$\langle p,\sigma;q,\varsigma|\hat{S}|\ell,\varepsilon_{\lambda}\rangle$},
one can finger out the dependence of the NBW pair production on the photon polarization and how the photon polarization transfers to the spin polarization of the produced particles.
Below, the deviation of the scattering matrix element in planed wave backgrounds is presented.


\subsection{Polarized NBW in pulsed plane waves}\label{Sec_Single_photon_pair_creation}
In modern-day laser-particle experiments~\cite{slacref1,Abramowicz:2021zja}, the NBW pair production is measured with the typical scenario of the collision between a beam of high-energy seed photons and a pulsed laser background which is simplified as a plane wave with scaled vector potential $a^{\mu}=|e|A^{\mu}(\phi)$ and wavevector $k^\mu = \omega (1,0,0,-1)$, where $\phi=k\cdot x$, $\omega$ is the laser frequency and $|e|$ is the charge of the positron.
The collisions happen within a small angle $\theta \ll 1$ close to the head-on geometry.
This planed background should be a good approximation for collisions between high-energy particles with weakly-focussed pulses~\cite{DiPiazza2015PRA,DiPiazza2016PRL,DiPiazza2017PRA032121,DiPiazza2021PRD076011}.

The scattering matrix element can now be written out explicitly,
\begin{align}
S_{p\sigma;q\varsigma;\ell\lambda}&=\langle p,\sigma;q,\varsigma|\hat{S}|\ell,\varepsilon_{\lambda}\rangle\nonumber\\
                                  &=-ie\int \ud^4x \overline{\varPsi}_{p,\sigma}(x)\slashed{A}_{\trm{ph}} \varPsi^{+}_{q,\varsigma}(x)\,,
\end{align}
where $\varPsi^{\LCp}_{q,\varsigma}$($\varPsi_{p,\sigma}$) is the Volkov wave function of the produced positron (electron) with the momentum $q^{\mu}$ ($p^{\mu}$) and the spin $\varsigma/2$ ($\sigma/2$)~\cite{wolkow1935klasse};
 \begin{align}
\varPsi_{p,\sigma}(x)&=\sqrt{\frac{m}{Vp^{0}}}\left(1-\frac{\slashed{k}\slashed{a}}{2k\cdot p}\right)u_{p,\sigma}e^{-ip\cdot x+i\int^{\phi}\ud\phi'
\frac{2p\cdot a+a^2}{2k\cdot p} },\nonumber\\
\varPsi^{\LCp}_{q,\varsigma}(x)&=\sqrt{\frac{m}{Vq^{0}}}\left(1+\frac{\slashed{k}\slashed{a}}{2k\cdot q}\right)v_{q,\varsigma}e^{iq\cdot x+i\int^{\phi}\ud\phi'\frac{2q\cdot a-a^2}{2k\cdot q} }.\nonumber
\end{align}
The explicit expression of the bispinor $u_{p,\sigma}$ ($v_{q,\varsigma}$) is given in the App.~\ref{Definition_of_Bispinors} in terms of the lightfront spin quantization axis
\begin{align}
S^{\mu}_{p}&=\frac{p^{\mu}}{m}-\frac{m}{k\cdot p}k^{\mu}\,,
\label{Eq_lightfront_spin_quantization}
\end{align}
($p^{\mu}$ is replaced with $q^{\mu}$ for $v_{q,\varsigma}$.) which is antiparallel to the laser propagating direction in the particle rest frame.
A different spin quantization axis, oriented along the direction of the magnetic field in the particle rest frame~\cite{BenPRA2015,seipt18,SeiptPRA052805}, is selected to avoid the BMT-precession~\cite{BMTPRL} of the axis.
As particle's spin quantum state cannot be changed by propagation through a plane wave in the absence of loop corrections or emissions~\cite{PRD2020Tang},
one can ignore the flip of the particle spin state in the intermediate intensity regime $\xi\sim O(1)$ and define the spin quantization axis~(\ref{Eq_lightfront_spin_quantization}) with the particles' asymptotic momentum ($p^{\mu}$ and $q^{\mu}$) after leaving the pulse.
For ultra-high intensities $\xi\gg O(1)$, one may frame the spin quantization axis with the particles' local momentum after production.

$A_{\textrm{ph}}$ is the seed photon with the momentum $\ell^{\mu}$,
\begin{align}
A^{\mu}_{\textrm{ph}}=\sqrt{\frac{2\pi}{\ell^{0} V}}\varepsilon^{\mu}_{\lambda}e^{-i\ell\cdot x}\,,
\label{Eq_NBW_photon0}
\end{align}
where $\varepsilon_{\lambda}$ is the circularly polarized lightfront base
\begin{align}
\varepsilon^{\mu}_{\LCpm}=&\epsilon^{\mu}_{\LCpm}-\frac{\ell\cdot \epsilon_{\LCpm}}{k\cdot \ell}k^{\mu}\,,
\end{align}
satisfying $\ell\cdot \varepsilon_{\pm}=0$ and $k\cdot \varepsilon_{\pm}=0$, where $\epsilon_{\pm}=(\epsilon_{x}\pm i\epsilon_{y})/\sqrt{2}$ and $\epsilon_{x}=(0,1,0,0)$, $\epsilon_{y}=(0,0,1,0)$. The subscript `$\lambda=\pm$' marks the rotation direction of the state. 
The probability for the photon-photon scattering is negligible in the intermediate intensity regime~\cite{DinuPRD2014,king_heinzl_2016}, so that 
one may not expect appreciable deviation of the photon polarization before pair production in the upcoming laser-particle experiments.
In terms of the above polarization base, one can relate the polarization matrix $\rho_{\gamma}$ directly to the classical Stokes parameters~\cite{landau4} defined in App.~\ref{photon_pol_density}, as
\begin{align}
\rho_{\gamma}=\begin{bmatrix}
  \rho_{\gamma,++} & \rho_{\gamma,+-} \\
  \rho_{\gamma,-+} & \rho_{\gamma,--}
\end{bmatrix}=\frac{I}{2}\begin{bmatrix}
  1+\Gamma_{3} & \Gamma_{1}-i\Gamma_{2} \\
  \Gamma_{1}+i\Gamma_{2} & 1-\Gamma_{3}
\end{bmatrix}\,,
\end{align}
where $I=\langle \gamma;\ell,\varepsilon|\gamma;\ell,\varepsilon\rangle$ is the number of the beam photons (or the intensity of the photon beam) and is set to be unity in the following calculations.
The Stokes parameters ($\Gamma_{1}$, $\Gamma_{2}$, $\Gamma_{3}$) measure both the direction and degree of the photon polarization;
$\Gamma_{3}$ denotes the degree of the circular polarization, and $\Gamma_{1,2}$ describe the linear polarization degree.
The total polarization degree of the photon beam is given as \mbox{$D=\sqrt{\Gamma^{2}_{1}+\Gamma^{2}_{2}+\Gamma^{2}_{3}}$}
where $0\leq D\leq1$; $D=1$ means a completely polarized photon, and $D=0$ an unpolarized photon.

After simple derivation, one can get:
\begin{align}
S_{p\sigma;q\varsigma;\ell\lambda}&=\frac{-ie}{k^{0}}\sqrt{\frac{2\pi m^{2}}{V^{3}q^{0}p^{0} \ell^{0}}}(2\pi)^{3}\delta^{\LCperp,\LCp}(p+q-\ell)\nonumber\\
                                  &~\int \ud \phi~\bar{u}_{p,\sigma}M(\varepsilon_{\lambda},\phi)v_{q,\varsigma} ~ e^{i\int^{\phi}d\phi' \frac{\ell\cdot\pi_{q}(\phi')}{k\cdot p}}
\end{align}
where $\pi_{q}$ is the instantaneous momentum of the positron,
\begin{align}
\pi^{\mu}_{q}(\phi)=q^{\mu}-a^{\mu}(\phi)+\frac{q\cdot a(\phi)}{k\cdot q} k^{\mu} -\frac{a^2(\phi)}{2k\cdot q} k^{\mu}\,,
\end{align}
and
\begin{align}
M(\varepsilon_{\lambda},\phi)&=\slashed{\varepsilon}_{\lambda}+\frac{\slashed{\varepsilon}_{\lambda}\slashed{k}\slashed{a}(\phi)}{2k\cdot q} -\frac{\slashed{a}(\phi)\slashed{k}\slashed{\varepsilon}_{\lambda}}{2k\cdot p}\,.
\label{Eq_Creation_matrix0}
\end{align}
The lightfront $\delta$-function $\delta^{\LCperp,\LCp}(p+q-\ell)$ guarantees the energy-momentum conservation in the process. The final spin density matrix can then be written as
\begin{align}
\rho_{{\sscript f},\varsigma\varsigma';\sigma\sigma'}
&=\sum_{\lambda\lambda'}\rho_{\gamma,\sscript\lambda\lambda'}\frac{\alpha}{(2\pi\eta)^2}\int \frac{\ud s}{ts} \int\ud^{2} \bm{r} \iint \ud \phi_{1} \ud \phi_{2}\nonumber\\
&~~~~~~~e^{i\int_{\phi_{2}}^{\phi_{1}}d\phi'\frac{\ell\cdot\pi_{q}(\phi')}{m^{2}\eta t}}~ \textrm{T}_{\sigma\varsigma\lambda;\sigma'\varsigma'\lambda'}(\phi_1,\phi_{2})\,,
\end{align}
which is parameterised by the three components of the positron momentum: these are $s=k\cdot q/k\cdot \ell$, the fraction of the photon lightfront momentum taken by the positron, and \mbox{$\bm{r}=(r_{x},r_{y})$}, \mbox{$r_{x,y}= q_{x,y}/m-s\ell_{x,y}/m$}, positron momenta in the plane perpendicular to the laser propagating direction.
The fraction of the lightfront momentum taken by the electron is $t=1-s$, and $\eta =k\cdot \ell/m^{2}$ characterises the colliding energy between the seed photon and laser photon.
The shorthand of the trace term is
\begin{align}
\textrm{T}_{\sigma\varsigma\lambda;\sigma'\varsigma'\lambda'}=\bar{u}_{p,\sigma}M(\varepsilon_{\lambda},\phi_{1})v_{q,\varsigma}
 \bar{v}_{q,\varsigma'}\overline{M(\varepsilon_{\lambda'},\phi_{2})} u_{p,\sigma'}\nonumber
\label{Eq_Trace0}
\end{align}
and $\overline{M(\varepsilon_{\lambda'},\phi)}=\gamma^{0}M^{\dagger}(\varepsilon_{\lambda'},\phi)\gamma^{0}$. By tracing out all or part of the spin states, one can acquire the total probability from the polarized seed photon
\begin{align}
\trm{P}=&\hat{\digamma}
         \sum_{\sigma,\varsigma}\sum_{\lambda\lambda'}\rho_{\gamma,\sscript\lambda\lambda'}\textrm{T}_{\sigma\varsigma\lambda;\sigma\varsigma\lambda'}(\phi_1,\phi_{2})~ \,,
\end{align}
and the density matrix of the produced positron ($\rho^{\LCp}_{\vsigma\vsigma'}$) and electron ($\rho^{\LCm}_{\sigma\sigma'}$),
\begin{subequations}
\begin{align}
\rho^{\LCp}_{\vsigma\vsigma'}=&\frac{1}{\trm{P}}\hat{\digamma}
                               \sum_{\lambda\lambda'}\rho_{\gamma,\sscript\lambda\lambda'} \sum_{\sigma} \textrm{T}_{\sigma\varsigma\lambda;\sigma\varsigma'\lambda'}(\phi_1,\phi_{2})~ \,,\\
\rho^{\LCm}_{\sigma\sigma'}  =&\frac{1}{\trm{P}}\hat{\digamma}
                               \sum_{\lambda\lambda'}\rho_{\gamma,\sscript\lambda\lambda'} \sum_{\vsigma} \textrm{T}_{\sigma\varsigma\lambda;\sigma'\varsigma\lambda'}(\phi_1,\phi_{2})~ \,,
\end{align}
\label{Eq_pair_spin_density_planed}
\end{subequations}
where the operator $\hat{\digamma}$ denotes the pre-integrals
\[
\hat{\digamma}\equiv \frac{\alpha}{(2\pi\eta)^2}\int \frac{\ud s}{ts}\int \ud^{2} \bm{r}\iint \ud \phi_{1}\ud \phi_{2}~ e^{i\int_{\phi_{2}}^{\phi_{1}}\ud\phi'\frac{\ell\cdot\pi_{q}(\phi')}{m^{2}\eta t}}.
\]

According to the spin density matrices~(\ref{Eq_pair_spin_density_planed}), one can easily acquire the spin polarization vectors $\bm{\Xi}_{\LCpm}$ of the produced electron and positron, as
\begin{align}
\rho^{\LCpm}=&\frac{1}{2}\left(\mathbb{1}_{2\times2}+\hat{\bm{\sigma}}\bm{\Xi}_{\LCpm}\right)\,,
\label{Eq_density_matrix}
\end{align}
where $\hat{\bm{\sigma}}=(\hat{\sigma}_{1},\hat{\sigma}_{2},\hat{\sigma}_{3})$ are the Pauli matrices.
\mbox{$\bm{\Xi}_{\LCpm}=(\Xi_{\LCpm,x},\Xi_{\LCpm,y},\Xi_{\LCpm,z})$} denotes the spin polarization degree in each direction;
$|\bm{\Xi}_{\LCpm}|=1$ corresponds to a pure state polarized in the direction $\bm{\Xi}_{\LCpm}$,
and $|\bm{\Xi}_{\LCpm}|<1$ denotes a mixed state polarized in the direction $\bm{\Xi}_{\LCpm}/|\bm{\Xi}_{\LCpm}|$ with the degree $|\bm{\Xi}_{\LCpm}|$. 

To calculate the trace terms\footnote{The trace calculations can be done using FEYNCALC~\cite{FeynCalc1,FeynCalc2}.}, one would need the outer products of the Dirac bispinors~\cite{DIEHL200341,PRD016005};
\begin{subequations}
\begin{align}
u_{p,\sigma'}\bar{u}_{p,\sigma}  &=\frac{\slashed{p}+m}{4m}\left(\delta_{\sigma\sigma'}+\gamma^{5}\slashed{S}_{p,\sigma\sigma'}\right)\,,\\
v_{q,\vsigma}\bar{v}_{q,\vsigma'}&=\frac{\slashed{q}-m}{4m}\left(\delta_{\vsigma\vsigma'}+\gamma^{5}\slashed{S}_{q,\vsigma\vsigma'}\right)\,,
\end{align}
\end{subequations}
where $\gamma^{5}=i\gamma^{0}\gamma^{1}\gamma^{2}\gamma^{3}$, and $S^{\mu}_{p,\sigma\sigma'}$ ($S^{\mu}_{q,\vsigma\vsigma'}$) is the covariant spin base vector according to the spin quantization axis~(\ref{Eq_lightfront_spin_quantization}), 
satisfying $p\cdot S_{p,\sigma\sigma'}=0$ ($q\cdot S_{q,\vsigma\vsigma'}=0$).
The expressions of these spin bases can be found in \mbox{App.~\ref{Definition_of_Bispinors}}.

\section{Spectrum and spin polarization}\label{Sec_3}
After calculating the Dirac trace, one can get the final probability of the NBW process from a polarized photon,
\begin{align}
\trm{P}=\hat{\digamma}
&\left\{h_s \bm{\Delta}^{2}/2 + 1 -i\Gamma_{3}h_s\bm{w}(\phi_{1})\times\bm{w}(\phi_{2})\right.\nonumber\\
&      -\Gamma_{1} \left[w_{x}(\phi_{1}) w_{x}(\phi_{2}) - w_{y}(\phi_{1}) w_{y}(\phi_{2})\right]\nonumber\\
&\left.-\Gamma_{2} \left[w_{x}(\phi_{1}) w_{y}(\phi_{2}) + w_{y}(\phi_{1}) w_{x}(\phi_{2})\right]\right\}\,,
\label{Eq_prob_polar0}
\end{align}
and the spin polarization vector of the produced positron,
\begin{widetext}
\begin{align}
\begin{bmatrix}
 \Xi_{\LCp,x} \\
 \Xi_{\LCp,y} \\
 \Xi_{\LCp,z}
\end{bmatrix} =
\frac{\hat{\digamma}}{\trm{P}}\begin{bmatrix}
~~\Delta_{y}/(2 s) \\
-\Delta_{x}/(2 s) \\
 -i g_s \bm{w}(\phi_{1})\times\bm{w}(\phi_{2})
\end{bmatrix} +
\frac{\hat{\digamma}}{\trm{P}}\begin{bmatrix}
 \Delta_{y}/(2t) & - \Delta_{x}/(2t) & -\Sigma_{x}/(2 s) \\
 \Delta_{x}/(2t) & ~~\Delta_{y}/(2t) & -\Sigma_{y}/(2 s) \\
 0 & 0 & 1/s + \bm{\Delta}^{2}g_s/2
\end{bmatrix} \begin{bmatrix}
 \Gamma_{1} \\
 \Gamma_{2} \\
 \Gamma_{3}
\end{bmatrix}\,,
\label{Eq_positron_dmatrix}
\end{align}
where $h_{s}=(s^2+t^2)/(2st)$, $g_s=(s-t)/(2st)$, \mbox{$\bm{\Delta} =i[\bm{a}^{\LCperp}(\phi_{1}) - \bm{a}^{\LCperp}(\phi_{2})]/m$}, \mbox{$\bm{w}(\phi)= \bm{r} - \bm{a}^{\LCperp}(\phi)/m$}, $\bm{w}(\phi_{1})\times \bm{w}(\phi_{2}) = w_{x}(\phi_{1})w_{y}(\phi_{2})-w_{y}(\phi_{1})w_{x}(\phi_{2})$, \mbox{$\bm{a}^{\LCperp}=(a_{x},a_{y})$}, and $\bm{\Sigma} = \bm{w}(\phi_{1}) + \bm{w}(\phi_{2})$. All the elements in the matrices would be operated by the integral operator $\hat{\digamma}$.

Integrating over the transverse momenta $\bm{r}$~\cite{DinuPRA2013}, one arrives at the lightfront momentum spectrum of the positron,
\begin{align}
\frac{\ud \trm{P}}{\ud s}=
&\hat{T}\left\{ h_{s}\bm{\Delta}^{2}/2 + 1- i\Gamma_{3}h_{s} \bm{\mathfrak{a}}(\phi_{1})\times \bm{\mathfrak{a}}(\phi_{2}) \right.\nonumber\\
&~~~~~~~~~~~~~~~~~~~~\left.-\Gamma_{1}\left[ \mathfrak{a}_{x}(\phi_{1}) \mathfrak{a}_{x}(\phi_{2})
                 - \mathfrak{a}_{y}(\phi_{1}) \mathfrak{a}_{y}(\phi_{2}) \right]
-\Gamma_{2}\left[ \mathfrak{a}_{x}(\phi_{1}) \mathfrak{a}_{y}(\phi_{2})
                       + \mathfrak{a}_{y}(\phi_{1}) \mathfrak{a}_{x}(\phi_{2}) \right]\right\}\,,
\label{Eq_prob_polar1_trans}
\end{align}
and the energy distribution of the positron's spin polarization vector,
\begin{align}
\begin{bmatrix}
 \Xi_{\LCp,x}(s) \\
 \Xi_{\LCp,y}(s) \\
 \Xi_{\LCp,z}(s)
\end{bmatrix} &=\frac{\hat{T}}{\ud\trm{P}/\ud s}
\begin{bmatrix}
~~\Delta_{y}/(2 s) \\
 -\Delta_{x}/(2 s) \\
-ig_{s}\bm{\mathfrak{a}}(\phi_{1})\times\bm{\mathfrak{a}}(\phi_{2})
\end{bmatrix} + \frac{\hat{T}}{\ud\trm{P}/\ud s}\begin{bmatrix}
 \Delta_{y}/(2t) & -\Delta_{x}/(2t) &- \mathfrak{b}_{x}/(2 s) \\
 \Delta_{x}/(2t) &~~\Delta_{y}/(2t) &- \mathfrak{b}_{y}/(2 s) \\
 0 & 0 & 1/s+g_{s}\bm{\Delta}^{2}/2
\end{bmatrix}\begin{bmatrix}
 \Gamma_{1} \\
 \Gamma_{2} \\
 \Gamma_{3}
\end{bmatrix}
\label{Eq_positron_dmatrix_trans}
\end{align}
\end{widetext}
which is normalized with the energy spectrum, where $\hat{T}$ is the integral operator given as
\[\hat{T}\equiv
\frac{i\alpha}{2\pi\eta}\int\ud\vphi \int\frac{\ud\vtheta}{\vtheta}~e^{\frac{i\vtheta\Lambda}{2\eta t s}}\,,\]
and $\bm{\mathfrak{a}}(\phi) = \left[\langle \bm{a}^{\LCperp} \rangle - \bm{a}^{\LCperp}(\phi)\right]/m$, $\bm{\mathfrak{b}}= \bm{\mathfrak{a}}(\phi_{1})+\bm{\mathfrak{a}}(\phi_{2}) $, the Kibble mass $\Lambda=1+\langle a\rangle^{2}/m^{2}-\langle a^2\rangle/m^2$~\cite{kibble64}, $\vphi=(\phi_{1}+\phi_{2})/2$ is the average phase and $\vtheta=\phi_{1}-\phi_{2}$ is the interference phase~\cite{dinu16,seipt2017volkov,king19a}. The window average $\langle f\rangle$ is calculated as $\langle f\rangle=\frac{1}{\vtheta}\int^{\vphi+\vtheta/2}_{\vphi-\vtheta/2}\ud \phi f(\phi)$.
The details of the calculation can refer to the analogous calculation presented in Ref.~\cite{BenPRA2020} for the polarized nonlinear Compton scattering.

As one can see in (\ref{Eq_prob_polar1_trans}), the yield of the process is comprised of the unpolarized contribution \hbox{$(\Gamma_{1,2,3}=0)$}~\cite{TangPRA2021,TangPRD096019}, and the contributions coupling to the photon polarization $(\Gamma_{1,2,3} \neq 0)$.
The linear polarization $\Gamma_{1,2}$ of the seed photon is coupled, respectively, to the preponderance of the field contribution in the $x$-direction over that in the $y$-direction,
and the preponderance of the field contribution in the direction with the azimuthal angle $\psi=45^{\circ}$ over that with $\psi=135^{\circ}$, as $\mathfrak{a}_{45^{\circ}}(\phi_{1}) \mathfrak{a}_{45^{\circ}}(\phi_{2}) - \mathfrak{a}_{135^{\circ}}(\phi_{1}) \mathfrak{a}_{135^{\circ}}(\phi_{2})$, where $\mathfrak{a}_{\psi}=\mathfrak{a}_{x}\cos\psi + \mathfrak{a}_{y}\sin\psi$.
The photon's circular polarization $\Gamma_{3}$ is related to the rotation property of the background fields, $\bm{\mathfrak{a}}(\phi_{1})\times\bm{\mathfrak{a}}(\phi_{2})$ which resulting from the phase delay between the fields in orthogonal  directions~\cite{jackson1999classical} and would be zero for linearly polarized laser backgrounds.

As one can see in (\ref{Eq_positron_dmatrix_trans}), the spin of the produced positron could not only be polarized by the background field, but also be transferred from the photon polarization.
The transverse \ST{spin component} $\Xi_{\LCp,x/y}$ is polarized directly by the magnetic field \hbox{($B_{x/y}\sim a'_{y/x} \sim \Delta_{y/x}$)} in the parallel direction with the polarization degree determined by the field asymmetry (between the positive and negative parts) in each direction~\cite{seipt18,Daniel2019PRA061402,2020PRA023102},
while the longitudinal \ST{component} $\Xi_{\LCp,z}$ is polarized by the rotation of the background field.
The transfer efficiency from the photon polarization to the positron spin polarization depends on the positron momentum and the form of the field.
As one can also see in (\ref{Eq_positron_dmatrix_trans}),
the transverse \ST{components $\Xi_{\LCp,x/y}$} relate to the laser field of the first order which would go to zero after the phase integral for a relatively long laser pulse with low field asymmetry (see the LMA discussion later).
Therefore, these transverse \ST{spin components are} unpolarized $\Xi_{\LCp, x/y}\approx0$ in this type of field and only the circular polarization of the seed photon, \emph{i.e.} the photon helicity, can transfer to the longitudinal \ST{polarization} (or helicity) of the positron, as which depends on the even order of field.
One should realize that the angular momentum conservation in this photon polarization transfer is generally fulfilled by the sum of the spin and orbital angular momentum of the produced electron and positron.

The corresponding results for the produced electron are presented in App.~\ref{Electron_result}. One can clearly see the symmetry between the produced positron and electron by exchanging their lightfront momenta $s\leftrightarrow t$ and considering their charge with different sign. The transverse integral over the positron's momentum can also be regarded as the integral over the transverse momentum of the electron following \mbox{$\bm{r}= \bm{q}^{\LCperp}/m-s\bm{\ell}^{\LCperp}/m = t\bm{\ell}^{\LCperp}/m - \bm{p}^{\LCperp}/m $}.

\subsection{Polarized locally monochromatic approximation}
The upcoming laser-particle experiments have been designed to concentrate on the intermediate intensity regime, where $\xi\sim O(1)$~\cite{slacref1,Abramowicz:2021zja},
in which the NBW process is regarded to happen in the length scale comparable to the laser wavelength~\cite{ritus85}
and thus the interference effect over the wavelength could play an importance role in the production process~\cite{ilderton2019coherent,Ilderton:2019vot,ILDERTON2020135410}.
To catch this interference effect in simulations, one can resort to the locally monochromatic approximation (LMA), \emph{e.g.} Ptarmigan \cite{ptarmigan}.
The LMA treats exactly the fast variation of the laser carrier frequency but neglects the slow variation of the pulse envelope~\cite{LMA063110,Blackburn_2021,blackburn2021higher}.
This request the pulse duration to be much longer than the laser period.
The shortcomings of the LMA have been discussed in~\cite{TangPRD096019}.

\subsubsection{Circularly polarized background}
For a circularly polarized laser background as:
\begin{align}
a^{\mu}(\phi)=m \xi~(0,\cos\phi, \mathfrak{c} \sin\phi,0)~f(\phi)\,,
\label{Eq_Com__plane_cir1_slow}
\end{align}
where $\mathfrak{c}=\pm1$ denotes the rotation direction of the field corresponding to the polarization state $(\epsilon_{x}+i\maf{c} \epsilon_{y})/\sqrt{2}$,
$\xi$ is the normalized pulse amplitude and $f(\phi)$ is the pulse envelope with the derivative $f'(\phi)\ll1$ describing the slow variation of the pulse amplitude,
the phase term in the integral operator $\hat{\digamma}$ in (\ref{Eq_prob_polar0}) can be written approximately as
\begin{align}
\int_{\phi_{2}}^{\phi_{1}}d\phi \frac{\ell\cdot\pi_{q}(\phi)}{k\cdot p}
\approx&\kappa(\vphi)\phi_{1} - \zeta(\vphi) \sin(\phi_{1} -\mathfrak{c}\psi )\nonumber\\
      -&\kappa(\vphi)\phi_{2} + \zeta(\vphi) \sin(\phi_{2} -\mathfrak{c}\psi )\,,\nonumber
\end{align}
where all the terms proportional to $f'(\phi)$ is neglected, \mbox{$\kappa(\vphi) = [1+r^{2} + \xi^{2}(\vphi)]/(2\eta t s)$}, \mbox{$\zeta(\vphi)= \xi(\vphi)r/(\eta t s)$}, and \mbox{$\xi(\vphi) = \xi f(\vphi)$} is the local intensity at the average phase $\vphi = (\phi_{1}+\phi_{2})/2$, and $r=|\bm{r}|$, \mbox{$\bm{r}=r(\cos\psi,\sin\psi)$}.

By neglecting the slow variation of the parameters $\kappa(\vphi)$ and $\zeta(\vphi)$, one can then do the Jacobi-Anger expansions for the fast-varying terms in the integrals over $\phi_{1,2}$~\cite{landau4}, and acquire
\begin{align}
\trm{P}&=\frac{\alpha}{2\pi \eta^{2}_{\ell}}\int \frac{\ud s}{ts}\int \ud^{2}\bm{r} \sum^{\infty}_{n=-\infty}  \int \ud\vphi~\delta[\kappa(\vphi) -n]\nonumber\\
       &\left\{ J^{2}_{n} + \xi^{2}(\vphi)\left[n^2 J^{2}_{n}/\zeta^2(\vphi) + J'^{2}_{n}- J^{2}_{n}\right]h_{s}\right.\nonumber\\
& + 2 \maf{c}\Gamma_{3}\xi(\vphi) h_{s}\left[r - \xi(\vphi) n/\zeta(\vphi) \right] J'_{n} J_{n}\nonumber\\
&       - \Gamma_{1}\left[(r - n\eta st/r )^{2}J^{2}_{n} - \xi^{2}(\vphi) J'^{2}_{n} \right]\cos2\psi\nonumber\\
&\left. - \Gamma_{2}\left[(r - n\eta st/r )^{2}J^{2}_{n} - \xi^{2}(\vphi) J'^{2}_{n} \right]\sin2\psi \right\}\,,
\label{Eq_NBW_polar_LMA_angle}
\end{align}
where the $\delta$-function comes from the integral over the interference phase $\vtheta=\phi_{1}-\phi_{2}$, $J_{n}\equiv J_{n}[\zeta(\vphi)]$ is the Bessel function of the first kind. The two harmonic sums [\emph{i.e.}, $\sum_{n_{1}}(\cdots)$ and $\sum_{n_{2}}(\cdots)$] from the Jacobi-Anger expansions are reduced to a single one by considering that the harmonic phase, $\exp[-i(n_{1}-n_{2})\vphi]$, supports the final probability mainly when $n_{1}=n_{2}=n$~\cite{nikishov64},
otherwise, the rapid oscillation would be induced into the phase integral in (\ref{Eq_NBW_polar_LMA_angle}). (One should note that for circularly polarized backgrounds, the reduction of the harmonic sums can also be done by integrating the azimuthal angle: $\int \ud\psi~\exp[i\mathfrak{c}(n_{1}-n_{2})\psi]=2\pi\delta_{n_{1},n_{2}}$~\cite{LMA063110}.)

As one can see from (\ref{Eq_NBW_polar_LMA_angle}),
the linear polarization ($\Gamma_{1}$ and $\Gamma_{2}$) of the seed photon, relating to the factors $\cos2\psi$ and $\sin2\psi$, could break the azimuthal symmetry of the positron distribution in circularly polarized laser backgrounds,
but not affect the yield of the produced particles as $\int \ud\psi (\cos2\psi, \sin2\psi) =(0,0)$. After integrating the transverse momentum, one can arrive at the positron spectrum:
\begin{align}
\frac{\ud\trm{P}}{\ud s} =&\frac{\alpha}{ \eta}  \int \ud\vphi \sum_{n=\lceil n_{\ast} \rceil}\left[J^{2}_{n} + \xi^{2}\left(\frac{n^2}{\zeta^2_{n}}J^{2}_{n} + J'^{2}_{n}-J^{2}_{n}\right)h_{s}\right.\nonumber\\
                          &\left.~~~~~~~~+2\maf{c}\Gamma_{3}\xi(\vphi) h_{s}\left(r_{n} - \frac{n\eta t s}{r_{n}} \right) J'_{n} J_{n}\right]\,,
\label{Eq_NBW_polar_LMA_S}
\end{align}
where $\lceil n_{\ast} \rceil$ denotes the lowest integer greater than or equal to $n_{\ast}=[1+\xi^{2}(\vphi)]/(2\eta t s)$, the argument of the Bessel function becomes \mbox{$\zeta_n(\vphi)=\xi(\vphi)r_{n}/(\eta t s)$}, and \mbox{$r_{n} =\sqrt{2n \eta t s -1 - \xi^{2}(\vphi)}$}.


In the same procedure, one can acquire the polarization vector of the positron $\bm{\Xi}_{\LCp}(s)$:
\begin{subequations}
\begin{align}
&\Xi_{\LCp,x}(s)=0\,,\\
&\Xi_{\LCp,y}(s)=0\,,\\
&\Xi_{\LCp,z}(s)=\frac{\alpha/\eta }{\ud\trm{P}/\ud s}\int \ud\vphi \sum_{n=\lceil n_{\ast} \rceil}\\
               &~~~~~~\left\{2 \maf{c}~\xi(\vphi) g_{s}~J_{n} J'_{n} ~\left(r_{n} - n\eta t s / r_{n}\right)\right.\nonumber\\
               &~~~~~~\left.+\Gamma_{3}\left[J^2_{n}/s + g_{s}\xi^2(\vphi)\left(n^2 J^2_{n}/\zeta^2_{n} + J'^2_{n} - J^2_n\right)\right]\right\}.\nonumber
\end{align}
\label{Eq_NBW_polar_LMA_Cir_spin}
\end{subequations}
As one can see, the transverse \ST{spin components} of the produced particles are unpolarized because the field in the LMA is symmetric in each direction,
and the longitudinal polarization of the positron spin stems from the rotation $\maf{c}$ of the background field,
which is proportional to the field intensity $\xi$,
and is transferred from the lightfront circular polarization of the seed photon $\Gamma_{3}$.
One should also note that the longitudinal \ST{spin polarization} $\Xi_{+,z}$ from the background field is anti-symmetric at $s=1/2$ because of the factors $g_{s}\propto 2s-1$, and that from the photon polarization is asymmetric in the energy distribution because of the factor $1/s$ in~(\ref{Eq_NBW_polar_LMA_Cir_spin}c).
These properties would result in the asymmetric energy spectrum of the positron on the spin state $\vsigma=\pm1$.

\subsubsection{Linearly polarized background}
In a linearly polarized laser background,
\begin{align}
a^{\mu}(\phi)=m \xi~(0,\cos\phi, 0, 0)f(\phi)
\label{Eq_Com__plane_cir1_slow}
\end{align}
with also slowly varying envelope $f'(\phi)\ll1$, the similar derivation as the circular case can be applied with the only complication introduced by the fast variation of $a^{2}(\phi)\sim \cos^{2}\phi$, the phase in (\ref{Eq_prob_polar0}) can now be calculated approximately as
\begin{align}
\int_{\phi_{2}}^{\phi_{1}}d\phi \frac{\ell\cdot\pi_{q}}{k\cdot p} &\approx \kappa(\vphi) \phi_{1}  - \zeta(\vphi)\sin\phi_{1} + \beta(\vphi) \sin(2\phi_{1}) \nonumber\\
                                                                        &- \kappa(\vphi) \phi_{2}  + \zeta(\vphi)\sin\phi_{2} - \beta(\vphi) \sin(2\phi_{2}),\nonumber
\end{align}
where \mbox{$\zeta(\vphi)= \xi(\vphi)r_{x}/(\eta st)$} and \mbox{$\beta= \xi^{2}(\vphi)/(8\eta s t)$}. The parameter $\kappa$ is given as in the circular case, except that $\xi^{2}\to \xi^{2}/2$.

By again neglecting the slow variation of the parameters $\kappa(\vphi)$, $\zeta(\vphi)$ and $\beta(\vphi)$, one can do the decomposition for the fast variations in the integrals over $\phi_{1,2}$, as
\[\Lambda_{j,n}(\zeta,\beta)=\int_{-\pi}^{\pi}\frac{\ud\phi}{2\pi}\cos^{j}(\phi)e^{i\left[n\phi-\zeta\sin(\phi)+\beta\sin(2\phi)\right]}\,,\]
where $j=0,1,2$, $\Lambda_{j,n}(\zeta,\beta)$ is the generalized Bessel functions~\cite{Reiss1962,landau4} and can be written as the sum of products of the first kind of Bessel functions~\cite{Korsch14947,PRE026707}, and following the similar procedure as in the circular case, finally arrive at
\begin{align}
\trm{P}=&\frac{\alpha}{2\pi \eta^{2}_{\ell}}\int \frac{\ud s}{ts}\int \ud^{2}\bm{r} \int \ud\vphi ~ \sum_{n=-\infty}^{+\infty}
         \delta[\kappa(\vphi) -n] \nonumber\\
       &\left\{\xi^{2}(\vphi)\left[\Lambda^{2}_{1,n} -\Lambda_{0,n}\Lambda_{2,n}\right]h_{s} + \Lambda^{2}_{0,n}\right.\nonumber\\
       &      -\Gamma_{1}\left[(r_{x}\Lambda_{0,n} - \xi(\vphi) \Lambda_{1,n})^{2} - \Lambda^{2}_{0,n}r^{2}_{y}\right] \nonumber\\
       &\left.-\Gamma_{2}2\left[r_{x}\Lambda_{0,n} - \xi(\vphi) \Lambda_{1,n}\right]\Lambda_{0,n} r_{y}\right\}\,,
\end{align}
After integrating the transverse momentum $r$, one can then obtain the positron spectrum
\begin{align}
\frac{\ud \trm{P}}{\ud s}&=\frac{\alpha}{\eta} \int \ud\vphi  \int^{\pi}_{-\pi}\frac{\ud \psi}{2\pi} \sum_{n=\lceil n_{\ast} \rceil}\\
        &\left\{\xi^{2}\left(\Lambda^{2}_{1,n} -\Lambda_{0,n}\Lambda_{2,n}\right)h_{s}+ \Lambda^{2}_{0,n}\right.\nonumber\\
        &\left. -\Gamma_{1}\left[(\Lambda_{0,n}r_{n}\cos\psi - \xi(\vphi)\Lambda_{1,n})^{2} - \Lambda^{2}_{0,n}r^{2}_{n}\sin^{2}\psi\right] \right\}\,,\nonumber
\label{Eq_NBW_polar_LMA_Spec_trans}
\end{align}
and the classical polarization vector
\begin{subequations}
\begin{align}
\Xi_{\LCp,x}(s)=&0\,,\\
\Xi_{\LCp,y}(s)=&0\,,\\
\Xi_{\LCp,z}(s)=&\frac{\Gamma_{3}}{\ud \trm{P}/\ud s}\frac{\alpha}{\eta}  \int \ud\vphi  \int^{\pi}_{-\pi}\frac{\ud \psi}{2\pi} \sum_{n=\lceil n_{\ast} \rceil} \nonumber\\
                &\left[\Lambda^{2}_{0,n}/s + \xi^{2}(\vphi)g_s \left(\Lambda^{2}_{1,n}- \Lambda_{0,n} \Lambda_{2,n}\right)\right]\,,
\end{align}
\label{Eq_NBW_polar_LMA_Lin_spin}
\end{subequations}
where $n_{\ast}=[1+ \xi^{2}(\vphi)/2]/(2\eta t s)$, and one of the arguments of $\Lambda_{j,n}(\zeta,\beta)$ is given as {$\zeta= \xi(\vphi) r_{n}\cos\psi/(\eta st)$} and \mbox{$r_{n}  = \sqrt{2n\eta t s - 1 - \xi^{2}(\vphi)/2}$}.

As one can see, in this linearly polarized laser background, the positron yield depends on the linear polarization $\Gamma_{1}$ of the seed photon, the other linear polarization $\Gamma_{2}$ would only affect the distribution of the produced positron. (The linear polarization $\Gamma_{2}$ could affect the positron yield if, for example, one rotates the major axis of the laser polarization to along the azimuthal angle $\psi=45^{\circ}$.)
The circular polarization of the seed photon $\Gamma_{3}$, even though not affects the yield of the production, transfers directly to the longitudinal \ST{polarization} of the produced positron,
while the positron's transverse spin polarization is again zero because of the symmetric field in $x$-direction.

\subsection{Polarized locally constant field approximation}
The locally constant field approximation (LCFA) has been widely employed in numerical simulations~\cite{2011PRL035001,2011PRSTAB054401,TangPRA2014,2015PRE023305,TangPRA2019,WAN2020135120,Seipt_2021} to investigate QED effects in classical backgrounds with ultra-high intensities $\xi\gg O(1)$.
The prerequisite of the LCFA is that the formation length of the process is much shorter than the typical length of the field variation~\cite{ritus85}, and the interference effect between the different points in the field could be ignored.
The whole process is thus regarded as the sum of the interactions with a series of infinitesimally short intervals of constant (crossed) field.
The range of application of the LCFA has been broadly studied~\cite{Piazza2018PRA012134,blackburn2020radiation},
and lots of effort have been paid to improve its precision~\cite{BenPRA2013,king19a,PiazzaPRA2019,BenPRA042508}

The LCFA is performed by expanding the interference phase $\vtheta$ to the first order for the field and the second order for the Kibble mass, as
\begin{align}
a^{\mu}(\phi_{1})&=a^{\mu}(\varphi)+a'^{\mu}(\varphi)\vtheta/2\,,\nonumber\\
a^{\mu}(\phi_{2})&=a^{\mu}(\varphi)-a'^{\mu}(\varphi)\vtheta/2\,,\nonumber\\
\Lambda & =1 + \vtheta^2 \xi^2(\varphi) /12\,,\nonumber
\end{align}
where $a'^{\mu}(\varphi)=m(0,-\xi_{x},-\xi_{y},0)$, $\bm{\xi}=(\xi_{x},\xi_{y})$ is the normalized electric field. Inserting into (\ref{Eq_prob_polar1_trans}) and (\ref{Eq_positron_dmatrix_trans}), one can obtain the positron spectrum
\begin{widetext}
\begin{align}
\frac{\ud\trm{P}}{\ud s}=&\frac{\alpha}{\eta } \int \ud \vphi~\left[ \textrm{Ai}_1(\zeta)-\frac{2}{\zeta}\textrm{Ai}'(\zeta)\left(h_{s}-\Gamma_{1} \frac{\xi^2_{x} -\xi^2_{y}}{2\xi^2}-\Gamma_{2}\frac{\xi_{x}\xi_{y}}{\xi^2}\right) \right],
\label{Eq_LCFA_Spectrum}
\end{align}
and the energy distribution of the polarization vector
\begin{align}
\begin{bmatrix}
 \Xi_{\LCp,x}(s) \\
 \Xi_{\LCp,y}(s) \\
 \Xi_{\LCp,z}(s)
\end{bmatrix} &=\frac{\alpha/\eta }{\ud\trm{P}/\ud s} \int \ud\vphi\left(
\begin{bmatrix}
~~\xi_{y}/ |\xi| \\
 -\xi_{x}/ |\xi| \\
       0
\end{bmatrix}\frac{\textrm{Ai}(\zeta)}{s\sqrt{\zeta}} + \begin{bmatrix}
 \frac{\xi_{y}}{|\xi| t\sqrt{\zeta}} \textrm{Ai}(\zeta)& -\frac{\xi_{x}}{|\xi| t\sqrt{\zeta}} \textrm{Ai}(\zeta) & 0 \\
 \frac{\xi_{x}}{|\xi| t\sqrt{\zeta}} \textrm{Ai}(\zeta)&~~\frac{\xi_{y}}{|\xi| t\sqrt{\zeta}} \textrm{Ai}(\zeta) & 0 \\
 0 & 0 & \textrm{Ai}_1(\zeta)\frac{1}{s} -\frac{2}{\zeta}\textrm{Ai}'(\zeta)g_{s}
\end{bmatrix}\begin{bmatrix}
 \Gamma_{1} \\
 \Gamma_{2} \\
 \Gamma_{3}
\end{bmatrix}\right)\,,
\label{Eq_positron_dmatrix_trans_LCFA}
\end{align}
\end{widetext}
where $\zeta=(s t \eta |\xi|)^{-2/3}$.
These results can be simply generalized to those in~\cite{SeiptPRA052805} where the spin basis along the magnetic field of a linearly polarized laser and the linear polarization basis for the seed photon were used.
One should note that the positron yield in~(\ref{Eq_LCFA_Spectrum}) does not depend on the circular polarization degree of the seed photon $\Gamma_{3}$, which contributes to the yield in~(\ref{Eq_prob_polar1_trans}) by coupling to the property of circular polarization of the background field.
This is because the fields in the infinitesimally short intervals, the background is split into by the LCFA, are \emph{linearly} polarized in the direction varying with the local phase $\vphi$.
The whole process is actually the integral over the consecutive interactions with infinitesimally short interval of linearly polarized constant field.
Therefore, the LCFA cannot capture the rotation property of the background field.
Under this approximation, the longitudinal \ST{polarization $\Xi_{\LCp,z}$} of the positron is proportional to the degree of the photon circular polarization $\Gamma_{3}$ as the linearly polarized LMA result in (\ref{Eq_NBW_polar_LMA_Lin_spin}).

\section{Numerical result}~\label{Sec_4}
In this section, we present the numerical example of a head-on collision between a $16.5$ GeV photon and a laser pulse with the wavelength $\lambda=0.8\mu m$.
The colliding energy parameter is thus $\eta \approx0.196$.
The laser pulse has the envelope $f(\phi)=\cos^{2}(\phi/4N)$ for $|\phi|<2N\pi$ with $f(\phi)=0$ otherwise, where $N=6$ corresponding the pulse duration of $16~\trm{fs}$ (full width at half maximum).
This setup is motivated by the upcoming particle-laser experiment in LUXE.

\subsection{Circularly polarized backgrounds}~\label{Circularly}
\begin{figure}[t!!!!]
\center{\includegraphics[width=0.45\textwidth]{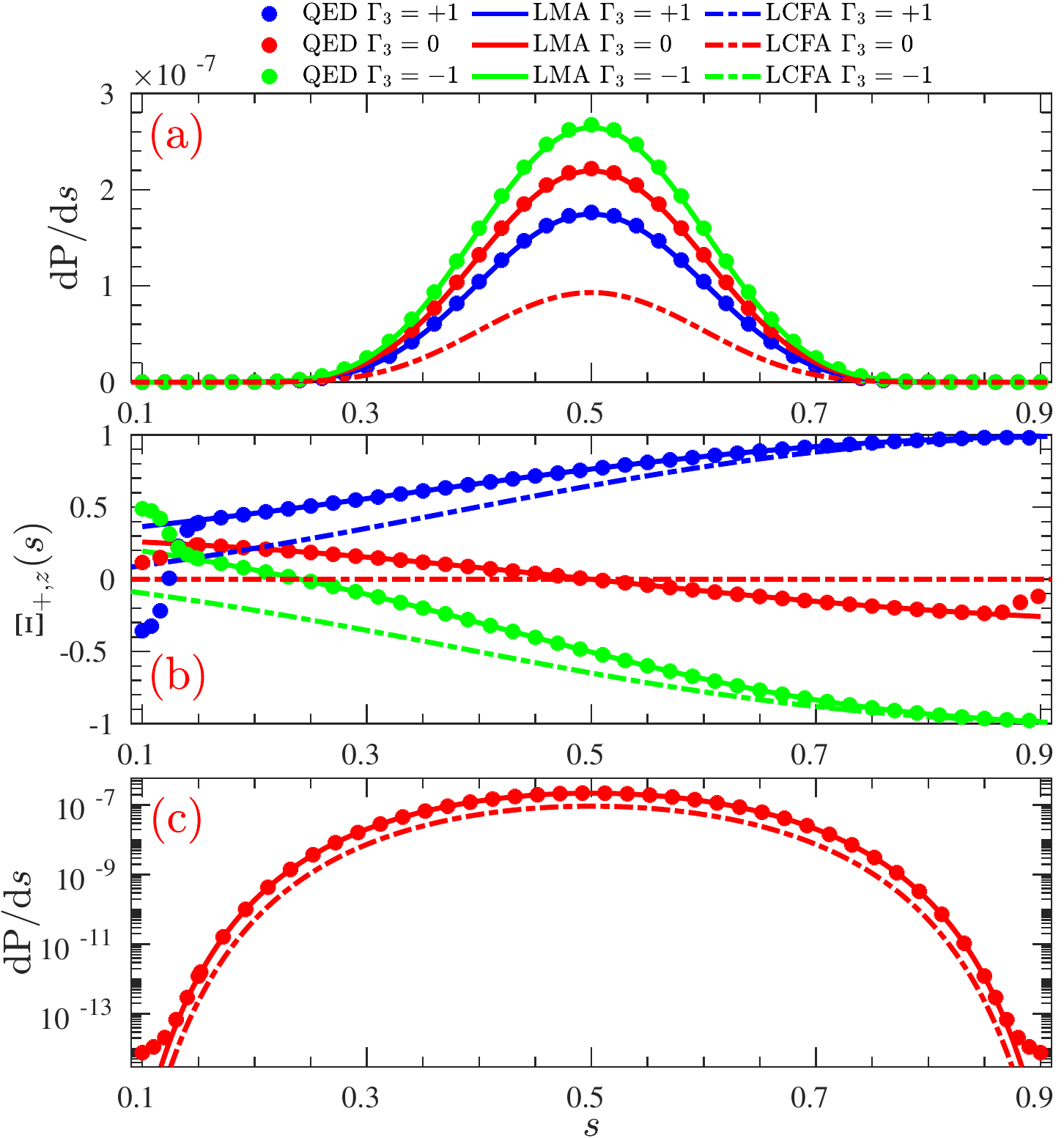}}
\caption{(a) Energy spectra of the produced positron in the NBW process stimulated by the seed photon with different circular polarization $\Gamma_{3}=0,~\pm1$.
(b) Distribution of the positron's longitudinal spin vector $\Xi_{\LCp, z}(s)$.
(c) Energy spectrum of the positron from the unpolarized photon plotted in the logarithmic scale.
The photon energy is $16.5~\trm{GeV}$.
The laser is circularly polarized with the intensity $\xi=1$ and carrier frequency $\omega=1.55~\trm{eV}$.
The pulse profile is $f(\phi)=\cos^{2}(\phi/4N)$ in the duration $|\phi|<2N\pi$, otherwise $f(\phi)=0$, where $N=6$.}
\label{Fig_Cir_Xi1}
\end{figure}

Figure~\ref{Fig_Cir_Xi1} presents the full QED calculations for the energy spectra and the longitudinal spin polarization of the positron produced via the \hbox{NBW} process in a circularly polarized laser pulse: \mbox{$a^{\mu}(\phi)=m \xi~(0,\cos\phi, \mathfrak{c} \sin\phi,0)~f(\phi)$}, where $\xi=1$ and $\maf{c}=1$, from a seed photon with different circular polarization degree $\Gamma_{3}=0,~\pm1$.
As shown in Fig.~\ref{Fig_Cir_Xi1} (a), the positron yield in this circularly polarized laser pulse depends substantially on the degree of the photon circular polarization; for the seed photon in the polarization state parallel to the background field, \emph{i.e.}, $\Gamma_{3}= \maf{c}$, the pair production (blue dotted) is suppressed with the lower spectrum than that (red dotted) from the unpolarized seed photon, while for the photon polarized in the state perpendicular to the background field, \emph{i.e.}, $\Gamma_{3}=-\maf{c}$, the pair production (green dotted) is enhanced.

The longitudinal \ST{polarization} $\Xi_{\LCp,z}(s)$ of the produced positron, shown in Fig.~\ref{Fig_Cir_Xi1} (b), depends also sensitively on the photon circular polarization $\Gamma_{3}$;
For an unpolarized photon $\Gamma_{3}=0$, the positron's longitudinal \ST{spin component} $\Xi_{\LCp,z}(s)$ is polarized by the rotation of the background field with a relatively small polarization degree, which is antisymmetric at $s=1/2$ because of the factor $g_{s}$ in (\ref{Eq_positron_dmatrix_trans}).
For a fully polarized photon $\Gamma_{3}=\pm1$, the spin polarization degree of the positron could be effectively improved, especially in the high-energy region ($s\to 1$),
in which the produced positron distributes in a narrow angular spread ($q_{\LCperp}\to 0$) collimated in the incident direction of the photon~\cite{TangPRA2021},
and thus the polarization angular momentum of the seed photon could transfer largely to the spin angular momentum of the produced particles.
While in the low-energy region ($s\to0$), the larger part of the photon's polarization angular momentum would be transferred to the orbital angular momentum ($\propto q_{\LCperp}$) of the produced particles because of their broader angular spread~\cite{TangPRA2021}.

The LMA results have been benchmarked with the full QED calculations in Fig.~\ref{Fig_Cir_Xi1}.
As shown in Fig.~\ref{Fig_Cir_Xi1} (a), the LMA can reproduce precisely the positron spectra from the differently polarized photons.
The only slight difference appears in the low- ($s\to0$) and high-energy ($s\to 1$) region where the \mbox{LMA} spectrum is lower than the full QED results, as shown in Fig.~\ref{Fig_Cir_Xi1} (c) for the unpolarized case.
This difference comes from the finite-pulse effect as the LMA approach actually describes the NBW process in an \emph{infinite} pulse with variable amplitude and thus ignores the effect of the photon entering and leaving the finite pulse~\cite{TangPRD096019}.
This effect also results in the discrepancy in the description of the positron's longitudinal spin polarization in the low- and high-energy region as shown in Fig.~\ref{Fig_Cir_Xi1} (b).
However, in the intermediate energy region ($0.2\lesssim s \lesssim 0.8$) covering the dominant production, the LMA describes the spin polarization of the produced positron very well.

The benchmarking of the LCFA results have also been presented in Fig.~\ref{Fig_Cir_Xi1}.
As shown in Fig.~\ref{Fig_Cir_Xi1} (a) for the positron spectra, the LCFA loses completely the contribution from the photon polarization $\Gamma_{3}$. This is because the LCFA treats the background field as a collection of instantaneous fields with linear polarization and thus loses the information of the field rotation, [see (\ref{Eq_LCFA_Spectrum})].
This shortcoming also results in the wrong prediction of the longitudinal polarization as shown in Fig.~\ref{Fig_Cir_Xi1} (b);
For an unpolarized photon, the LCFA predicts the produced positron with the unpolarized longitudinal \ST{spin component}, $\Xi_{\LCp,z}=0$, [see (\ref{Eq_positron_dmatrix_trans_LCFA})].
For a circularly polarized photon $\Gamma_{3}=\pm 1$, the LCFA results differ considerably from the full QED calculations in the lower-energy region ($s<0.5$) where the influence from the background field is crucial.
While with the increase of the positron energy ($s\to 1$), this difference becomes smaller as the polarization transferred from the seed photon becomes dominant.

\begin{figure}[t!!]
\center{\includegraphics[width=0.45\textwidth]{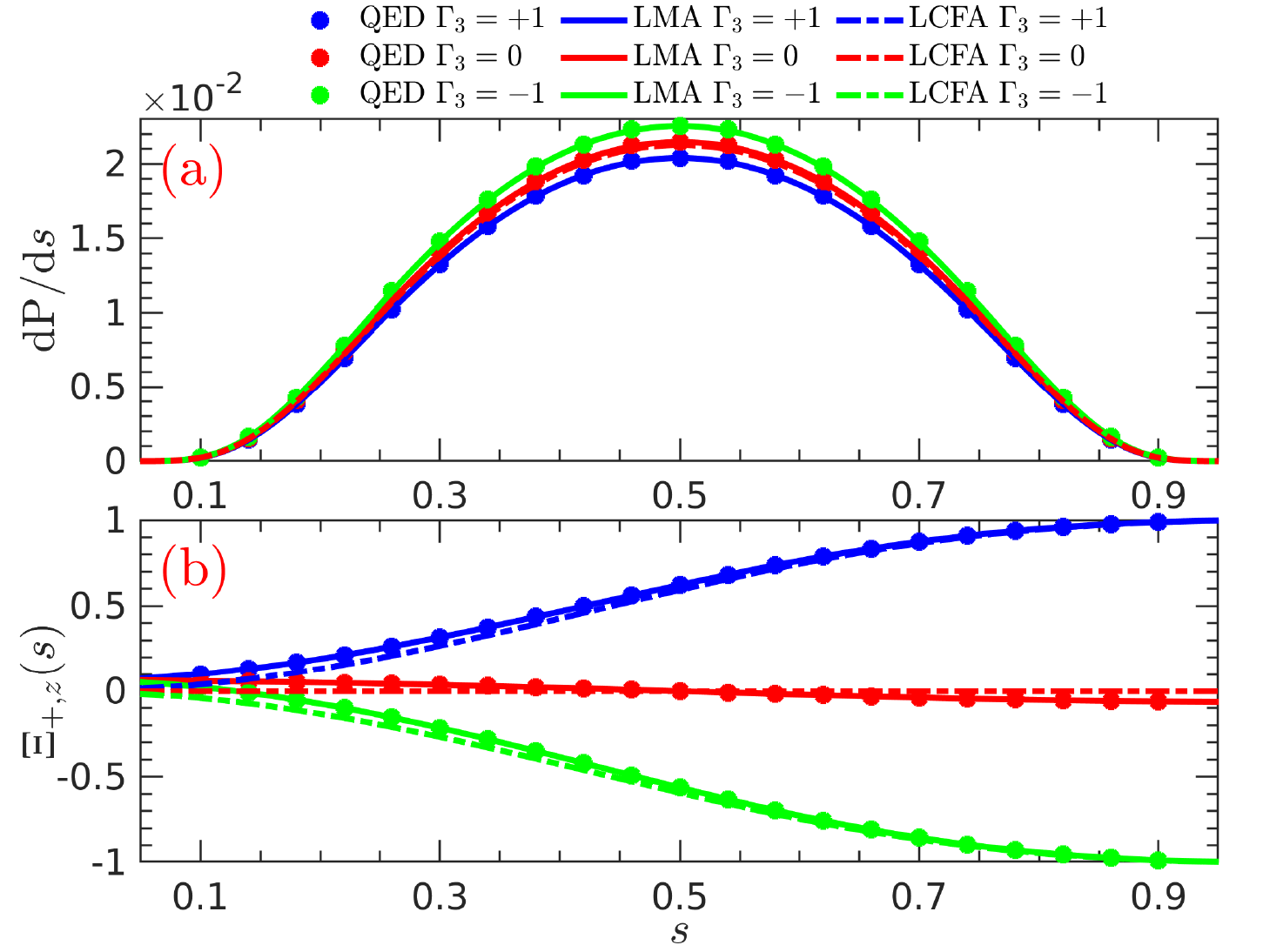}}
\caption{(a) Energy spectra of the produced positron in the NBW process stimulated by the seed photon with differen circular polarization $\Gamma_{3}=0,~\pm1$.
(b) Distribution of the positron's longitudinal \ST{polarization} $\Xi_{\LCp,z}(s)$.
All the parameters are the same as in Fig.~\ref{Fig_Cir_Xi1} except the laser intensity $\xi=5$ .}
\label{Fig_Cir_Xi5}
\end{figure}

The other shortcoming of the LCFA is the underestimation of the positron yield~\cite{BenPRA042508} as shown in Figs.~\ref{Fig_Cir_Xi1} (a) and (c). However, this underestimation becomes negligible when the laser intensity increases to $\xi=5$ in Fig.~\ref{Fig_Cir_Xi5}.
As shown in Fig.~\ref{Fig_Cir_Xi5} (a), the LCFA spectrum matches the full QED calculation very well for the unpolarized photon.
This improvement of the LCFA precision is because of the rapid decrease of the formation length of the \hbox{NBW} process with the increase of the pulse intensity~\cite{ritus85}.
As one can also see in Fig.~\ref{Fig_Cir_Xi5} (a), with the increase of the laser intensity, the importance\footnote{The ratio between the contribution from the photon polarization ($\Gamma_{3}$ for the circular case and $\Gamma_{1}$ for the linear case) and that from an unpolarized photon.} of the photon circular polarization for the positron yield decrease to about $5.3\%$ from about $20.9\%$ at $\xi=1$ in Fig.~\ref{Fig_Cir_Xi1} (a). Therefore, the \mbox{LCFA} result, at the high intensity $\xi\gtrsim5$, could be used to predict the positron yield from both polarized and unpolarized seed photon.
Simultaneously, the discrepancy in the positron's longitudinal \ST{polarization $\Xi_{\LCp,z}$} between the LCFA result and QED calculation becomes smaller,
as the spin polarized by the background field tends to zero, see the red lines in Fig~\ref{Fig_Cir_Xi5} (b).
The obvious difference appears only in the low-energy region where the spin polarization from the background is crucial as shown in Fig.~\ref{Fig_Cir_Xi5} (b).
With the increase of laser intensity, the finite-pulse effect becomes less important~\cite{TangPRD096019}, and can only result in the discrepancy in the extremely low- and high-energy region.

One should realize that for this circularly polarized laser pulse with relatively long duration, the influence of the photon's linear polarization $\Gamma_{1,2}$ is orders of magnitude smaller than that from the photon's circular polarization $\Gamma_{3}$, see (\ref{Eq_NBW_polar_LMA_S}),
while the importance of the linear polarization $\Gamma_{1}$ would increase significantly with the decrease of the pulse duration~\cite{TangPRD096019}.
The particle's transverse \ST{spin components} are unpolarized, $\Xi_{x,y}\approx 0$, because of the low-degree of the field asymmetry. For an ultra-short laser pulse or the laser pulse with effective unipolarity, the positron beam with high degree of transverse spin polarization can be produced~\cite{2020PRA023102,PRD2020Tang}.

\subsection{Linearly polarized backgrounds}~\label{Linearly}
\begin{figure}[t!!!!]
\center{\includegraphics[width=0.45\textwidth]{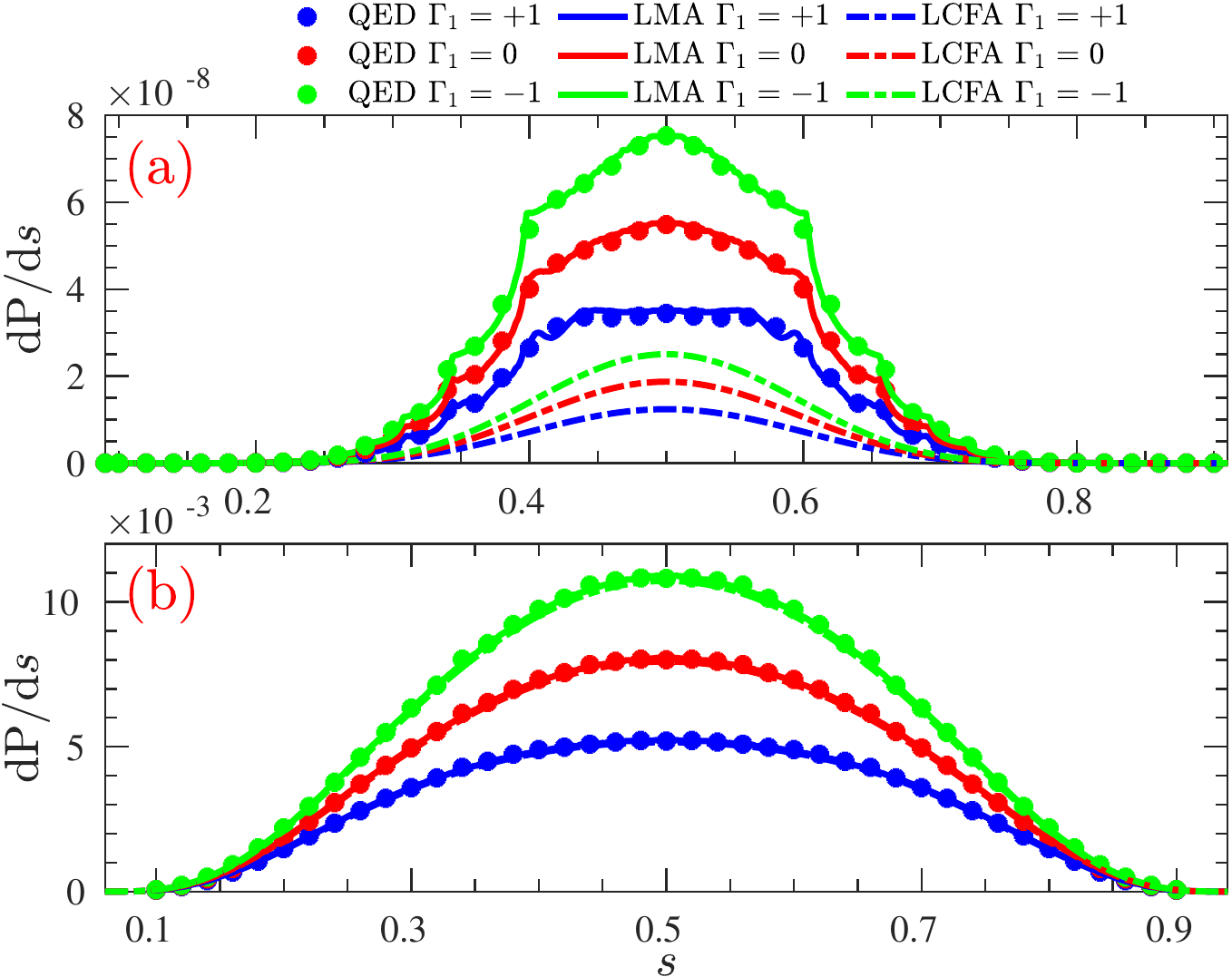}}
\caption{Energy spectra of the produced positron in the NBW process stimulated by a seed photon in the linear polarization state $\Gamma_{1}=0,\pm1$ in the laser background with the linear polarization and the intensity (a) $\xi=1$ (b) $\xi=5$. The other parameters are the same as in Fig.~\ref{Fig_Cir_Xi1}.}
\label{Fig_Lin_Spec}
\end{figure}
The energy spectra of the produced positron via the NBW process in a linearly polarized laser pulse \mbox{$a^{\mu}(\phi)=m \xi~(0,\cos\phi, 0, 0)~f(\phi)$}, are presented in Fig.~\ref{Fig_Lin_Spec} (a) for the intensity $\xi=1$ and (b) for $\xi=5$.
The seed photon is linearly polarized with the degree $\Gamma_{1}=0,\pm1$.
Similar as in the circular case, the yield of the production could be substantially enhanced when the seed photon is polarized in the state perpendicular to the laser polarization, \emph{i.e.}, $\Gamma_{1}=-1$, and be suppressed when the photon's polarization state is parallel to the laser pulse, \emph{i.e.}, $\Gamma_{1}=+1$.
With the increase of the laser intensity, the importance of the photon polarization would slightly decrease from about $33.0\%$ at $\xi=1$ in Fig.~\ref{Fig_Lin_Spec} (a) to about $30.6\%$ at $\xi=5$ in Fig.~\ref{Fig_Lin_Spec} (b).
This property suggests that the influence of the photon polarization on the positron yield could extend to the higher intensity region in the linearly polarized laser background than that in the background with circular polarization.
As shown in (\ref{Eq_NBW_polar_LMA_Spec_trans}), for the laser pulse linearly polarized in the $x$-direction, the contributions from the photon polarization $\Gamma_{2,3}$ are zero.

\begin{figure}[t!!!!]
\center{\includegraphics[width=0.45\textwidth]{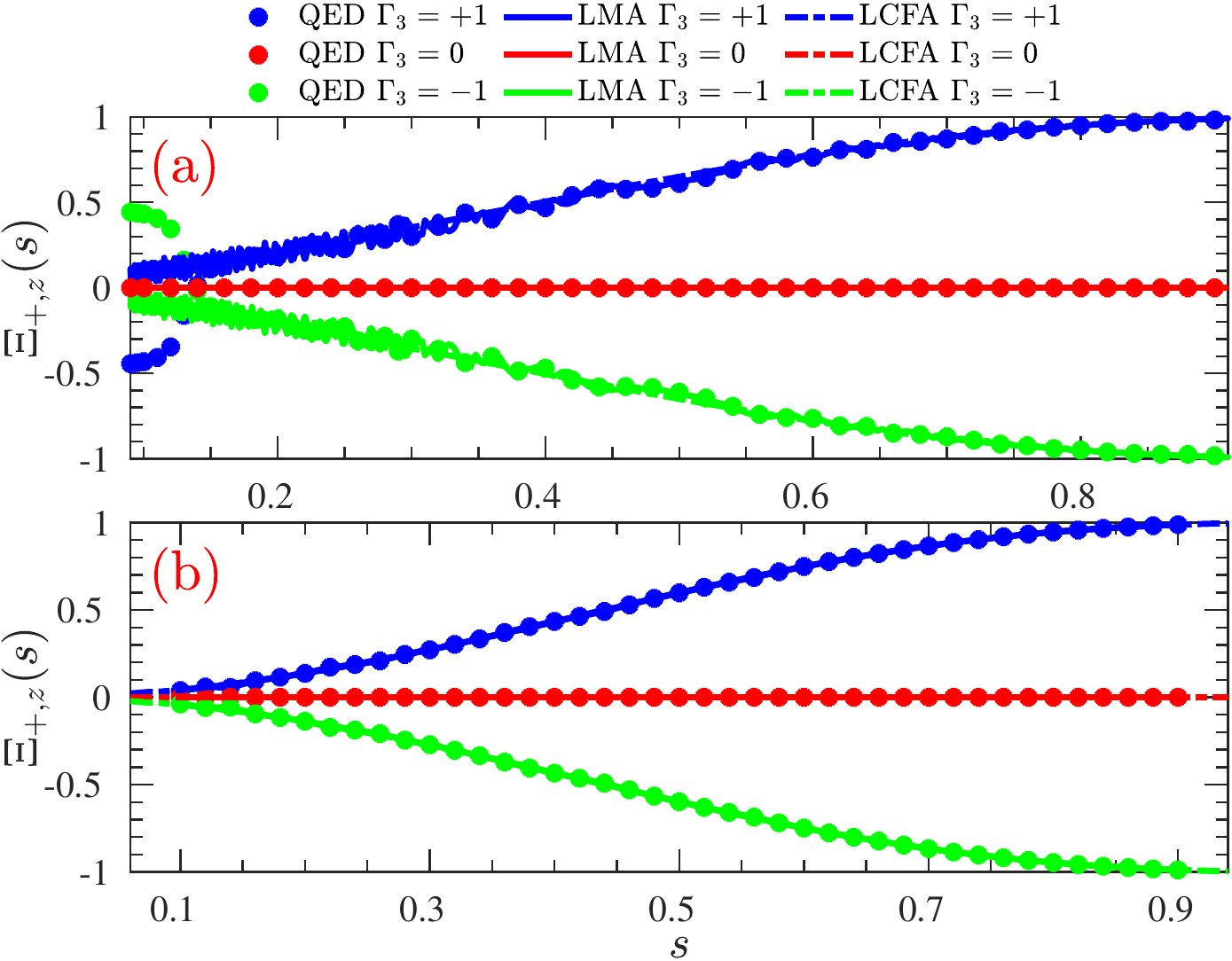}}
\caption{Distribution of the positron's longitudinal \ST{polarization degree} $\Xi_{\LCp,z}(s)$ produced via the NBW process stimulated by circularly polarized photons with the degree $\Gamma_{3}=0,\pm 1$ and $\Gamma_{1}=0$.
The laser background is linearly polarized with the intensity (a) $\xi=1$ and (b) $\xi=5$. The other parameters are the same as in Fig.~\ref{Fig_Cir_Xi1}.}
\label{Fig_Lin_Xiz}
\end{figure}

The spin polarization of the produced positron has been plotted in Fig.~\ref{Fig_Lin_Xiz}. 
As, again, for this long-duration laser pulse, the degree of the field asymmetry is significantly small, 
the transverse \ST{spin component} of the produced positron is unpolarized with $\Xi_{\LCp,x}\approx\Xi_{\LCp,y}\approx0$.
As one can see from (\ref{Eq_positron_dmatrix_trans}), 
the positron's longitudinal \ST{spin component $\Xi_{\LCp,z}$} can be polarized by the rotation of the background filed, 
which is also zero in linearly polarized pulses shown as the red lines in Fig.~\ref{Fig_Lin_Xiz} for $\Gamma_{3}=0$,
and can be transferred from the circular polarization of the seed photon shown as the green lines for $\Gamma_{3}=-1$ and blue lines for $\Gamma_{3}=+1$.
In the high-energy region, the polarization transfer to the positron spin is more efficient,
and highly polarized positron beam can be obtained with $s>0.5$.
One can also see the rapid oscillation in the positron longitudinal \ST{polarization $\Xi_{\LCp,z}$} because of the harmonic structure.

The LMA spectra have been benchmarked in Fig.~\ref{Fig_Lin_Spec}.
As shown, the LMA approach can not only reproduce the positron spectra from the differently polarized photon, but also describe precisely the harmonic structure in the positron spectra,
which appears at \mbox{$s_{n}=[1\pm \sqrt{1-(2+\xi^{2})/(n\eta )}]/2$} where $n$ is the integer greater than $n_{\trm{min}}=(2+\xi^{2})/\eta $.
Similar as the circular case, the slight discrepancy in the positron spectrum could only appear in the region far from the dominant spectral region because of the finite-pulse effect~\cite{TangPRD096019}, which would also result in the discrepancy in the LMA result for the positron's longitudinal \ST{polarization} at $s\to0$ shown in Fig.~\ref{Fig_Lin_Xiz} (a).
With the increase of the laser intensity, the finite-pulse effect becomes less important, and the discrepancy, between the LMA and full QED calculation, would only appear in the extremely low- and high-energy region.

For linearly polarized laser backgrounds, the LCFA can resolve the positron spectra from the seed photon with different linear polarization $\Gamma_{1}=0,\pm1$, but not describe the harmonic structures in the spectra as shown in \mbox{Fig.~\ref{Fig_Lin_Spec} (a)}.
Again, the LCFA underestimates the positron yield at the intensity $\xi=1$, and with the increase of the laser intensity, this underestimation becomes negligible as shown in Fig.~\ref{Fig_Lin_Spec} (b) for $\xi=5$.
The main tendency of the positron's spin polarization could also be captured by the LCFA except the slight harmonic fluctuations and that from the finite-pulse effect shown in Fig.~\ref{Fig_Lin_Xiz} (a).

\subsection{Polarization of the longitudinal spin vector}~\label{Longitudinal}
\begin{figure}[t!!!!]
\center{\includegraphics[width=0.35\textwidth]{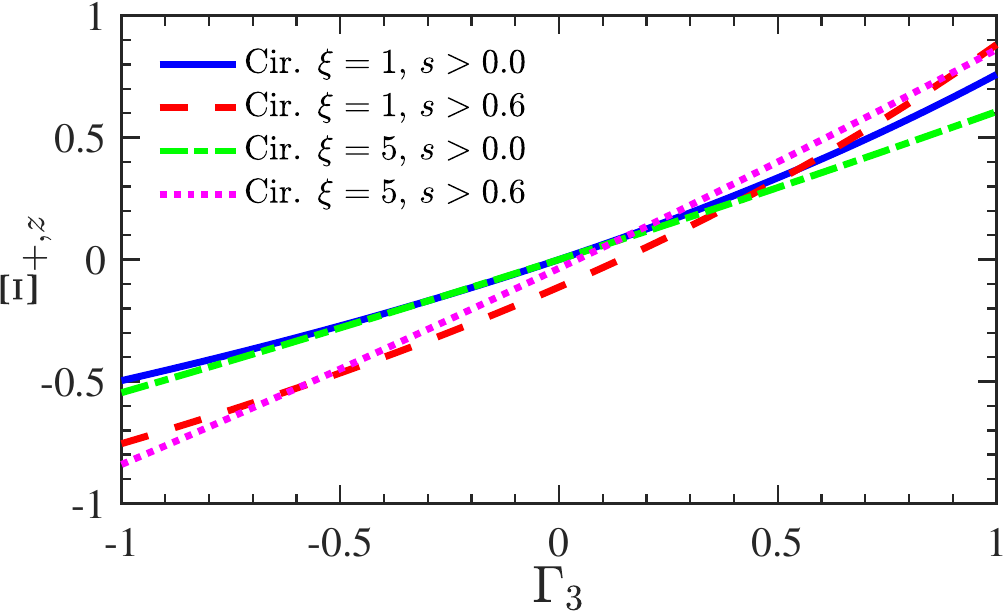}}
\caption{Longitudinal polarization degree $\Xi_{\LCp,z}$ of the positron beam with the change of the circular polarization degree of the incident photon beam $\Gamma_{3}$. The laser background is circularly polarized with the intensity $\xi=1$ (blue/solid line for $s>0$ and red/dashed line for $s>0.6$) and $\xi=5$ (green/dotted line for $s>0$ and magenta/dash-dotted line for $s>0.6$). The full QED calculation is applied with the same parameters in Fig.~\ref{Fig_Cir_Xi1}. $s=0.6$ corresponds to the positron energy $9.9~\trm{GeV}$.}
\label{Fig_Cir_Long_spin}
\end{figure}

The above numerical results [Fig.~\ref{Fig_Cir_Xi1}(b), Fig.~\ref{Fig_Cir_Xi5}(b) and Fig.~\ref{Fig_Lin_Xiz}] clearly suggest the generation of the positron (electron) beam with the highly polarized \ST{spin in the longitudinal direction} from a beam of circularly polarized seed photons with $\Gamma_{3}=\pm1$.
As one can see, in the circularly polarized laser background,
both the positron yield and its longitudinal \ST{polarization} depend on the circular polarization of the seed photon,
and both are free from the photon's linear polarization.
Therefore, the longitudinal polarization degree of the produced positron beam is, phenomenologically, the only function of the photon's circular polarization degree $\Gamma_{3}$, as
\begin{align}
\Xi_{\LCp,z}=\frac{\Xi_{z,0}+\Xi_{z,3} \Gamma_{3}}{\trm{P}_{0}+\trm{P}_{3} \Gamma_{3}}\,,
\label{Eq_longi_spin_cir}
\end{align}
where $\trm{P}_{0/i}$ ($\Xi_{z,0/i}$) are respectively the unpolarized and polarized contribution to the production yield (longitudinal \ST{polarization}),
and can be obtained from~(\ref{Eq_prob_polar1_trans}) [(\ref{Eq_positron_dmatrix_trans}) without the normalization by the spectrum $\ud\trm{P}/\ud s$] by integrating over the lightfront momentum $s$.
The subscript `$i$' means the contribution coupling to $\Gamma_{i}$.

\begin{figure}[t!!!!]
\center{\includegraphics[width=0.48\textwidth]{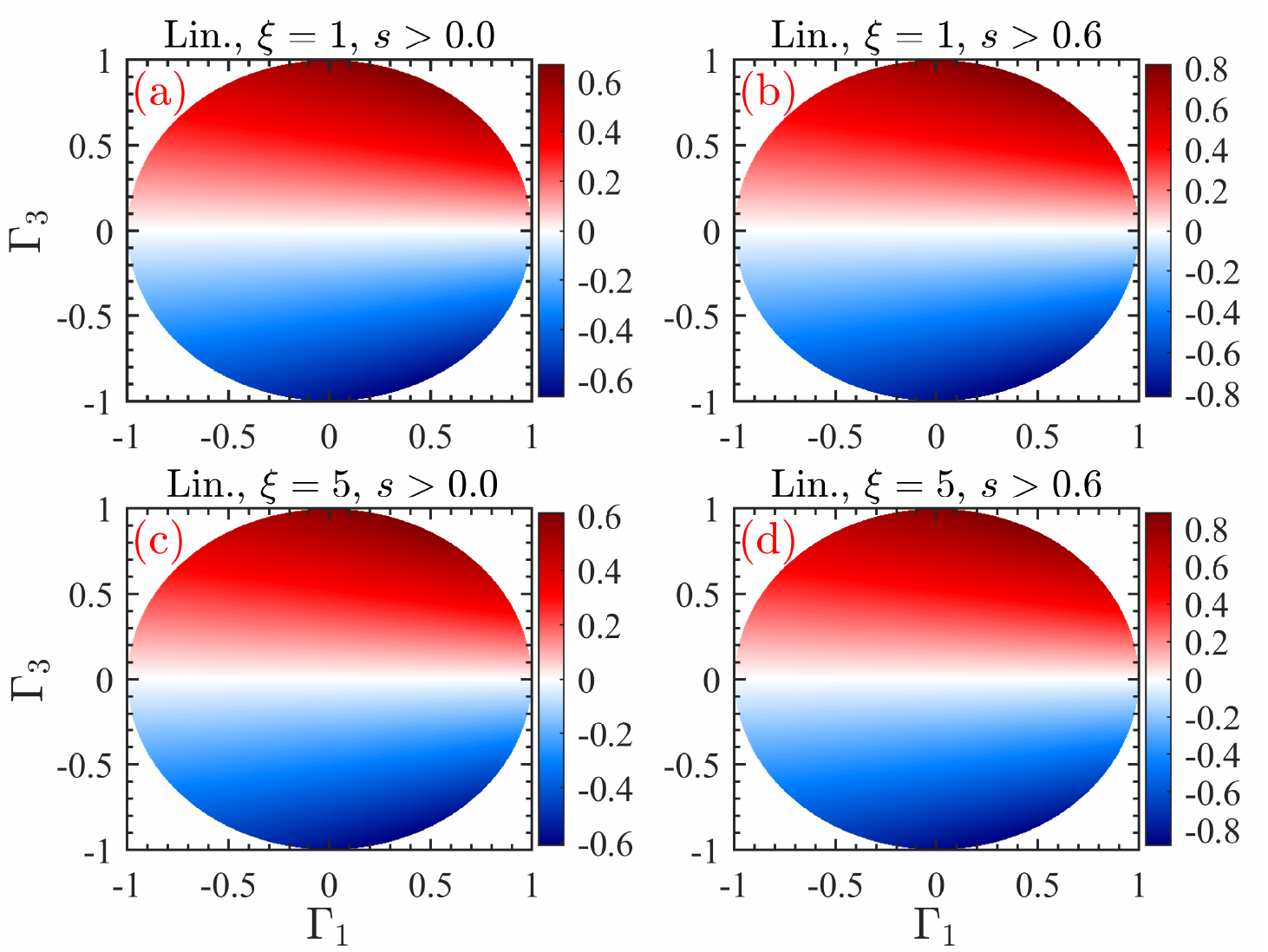}}
\caption{Longitudinal polarization degree $\Xi_{\LCp,z}$ of the produced positron varying with the polarization of the incident photon beam $(\Gamma_{1},~\Gamma_{3})$ where $\Gamma^{2}_{1}+\Gamma^{2}_{3}\leq 1$.
The laser background is linearly polarized with the intensity $\xi=1$ [(a) $s>0$ and (b) $s>0.6$] and $\xi=5$ [(c) $s>0$ and (d) $s>0.6$].
The full QED calculation is applied with the same parameters as in Fig.~\ref{Fig_Lin_Spec}.}
\label{Fig_Lin_Long_spin}
\end{figure}

Fig.~\ref{Fig_Cir_Long_spin} presents the dependence of the positron's longitudinal polarization $\Xi_{\LCp,z}$ on the circular polarization degree $\Gamma_{3}$ of the incident photon beam in circularly polarized laser fields.
As shown, if $\Gamma_{3}=0$, the positron beam is unpolarized as a whole ($s>0$) or slightly polarized for the high-energy part ($s>0.6$ corresponding to $9.9~\trm{GeV}$) as one can infer from Figs.~\ref{Fig_Cir_Xi1}(b) and~\ref{Fig_Cir_Xi5}(b).
However, with the increase of the photon's circular polarization degree $|\Gamma_{3}|\to1$,
the longitudinal polarization degree $\Xi_{\LCp,z}$ of the positron beam increases monotonously with the same sign as $\Gamma_{3}$,
to about $75.9\%$ at $\Gamma_{3}=+1$ and $-49.6\%$ at $\Gamma_{3}=-1$ for the whole positron beam ($s>0$) in the background with $\xi=1$, and
to about $\Xi_{\LCp,z}\approx 82.6\%,-63.8\%$ for its high-energy part with $s>0.6$ at $\Gamma_{3}=+1, -1$.
In the background with $\xi=5$, the polarization degree $\Xi_{\LCp,z}$ of the whole positron beam increases from about $-54.6\%$ at $\Gamma_{3}=-1$ to about $60.7\%$ at $\Gamma_{3}=+1$,
and for the high-energy positron beam with $s>0.6$, the longitudinal polarization degree increases from about $-76.0\%$ at $\Gamma_{3}=-1$ to about $79.2\%$ at $\Gamma_{3}=+1$.

In the linearly polarized laser background, the situation is analogous to the circular case with the only complication introduced as that the positron yield relies only on the photon's linear polarization $\Gamma_{1}$ while its longitudinal \ST{polarization degree} is proportional to the photon's circular polarization $\Gamma_{3}$, as
\begin{align}
\Xi_{\LCp,z}=\frac{\Xi_{z,3}~\Gamma_{3}}{\trm{P}_{0}+\trm{P}_{1} \Gamma_{1}}\,,
\label{Eq_longi_spin_lin}
\end{align}
where $\Gamma^{2}_{3}+\Gamma^{2}_{1}\leq 1$.
This property brings the antisymmetry into the spin polarization $\Xi_{\LCp,z}$ with \mbox{$\Gamma_{3}\in[-1,1]$},
and deviates the photon polarization, which supporting the highest longitudinal spin polarization, from the full circular polarization \mbox{($\Gamma_{1}=0, \Gamma_{3}=\pm1$)} in the circular case to the special polarization state \mbox{$(\Gamma_{1}, \Gamma_{3})=[-\trm{P}_{1}, \pm(\trm{P}^{2}_{0}-\trm{P}^2_{1})^{1/2}]/\trm{P}_{0}$}.
As shown in Fig.~\ref{Fig_Lin_Long_spin}, where the variation of the \ST{positron's longitudinal polarization} $\Xi_{\LCp,z}$ with the change of the photon polarization ($\Gamma_{1},\Gamma_{3}$) is presented,
the highest longitudinal \ST{polarization degree} of the whole positron beam in \mbox{Fig.~\ref{Fig_Lin_Long_spin} (a)} is about $\Xi_{\LCp,z}\approx \pm66.8\%$ coming from the photon beam with the polarization $(\Gamma_{1}\approx 0.33, \Gamma_{3}\approx\pm 0.94)$ in the linear background with $\xi=1$,
and for the high-energy positrons with $s>0.6$ in Fig.~\ref{Fig_Lin_Long_spin} (b),
the highest longitudinal polarization is close to $\pm81.7\%$ at \mbox{$(\Gamma_{1}\approx 0.31, \Gamma_{3}\approx\pm 0.95)$}.
In the background with $\xi=5$, the highest $\Xi_{\LCp,z}$ is about $\pm61.0\%$ for the whole positron beam and
about $\pm88.4\%$ for the positron beam with $s>0.6$ appearing respectively at $(\Gamma_{1}\approx 0.31, \Gamma_{3}\approx\pm 0.95)$ in Fig.~\ref{Fig_Lin_Long_spin} (c) and $(\Gamma_{1}\approx 0.28, \Gamma_{3}\approx\pm 0.96)$ in Fig.~\ref{Fig_Lin_Long_spin} (d).

\section{Conclusion}~\label{Sec_5}
Fully polarized nonlinear Breit-Wheeler pair production from a beam of polarized seed photons is investigated in pulsed plane-wave backgrounds.
Compact expressions for the energy spectrum and spin polarization of the produced electron and positron are derived depending on the photon polarization. Both the fermion spin and photon polarization are comprehensively described with the theory of the polarization density matrix.

The appreciable improvement of the positron yield is observed in the intermediate intensity regime \mbox{[$\xi\sim O(1)$]} when the photon polarization is orthogonal to the laser polarization;
For linearly polarized laser backgrounds, the yield could be improved about $30\%$ compared to the case from an unpolarized photon,
and for circularly polarized laser backgrounds, the improvement is about $21\%$ at the intensity $\xi=1$ and decreases to about $5\%$ at the intensity $\xi=5$.
The numerical results also suggest the generation of a high-energy positron beam with the highly polarized longitudinal \ST{spin component}, whose degree could be finely controlled by the circular polarization degree of the photon beam;
From a fully polarized photon beam with $\Gamma_{3}=\pm1$, a high-energy positron beam with the longitudinal polarization degree larger than $80\%$ could be produced in the intermediate intensity regime.

The locally monochromatic approximation and the locally constant field approximation are derived and benchmarked with the full QED calculations.
The locally monochromatic approximation can work precisely by reproducing not only the particle spectrum from the differently polarized photon, but also the spin polarization of the produced particles.
The slight discrepancy, originating from the finite-pulse effect, can only appear in the energy region far from the region with the dominant production.
However, the locally constant field approximation can only work well in the high-intensity regime ($\xi\gg1$).
In the intermediate intensity regime, which would be the mainly explored in the upcoming laser-particle experiments,
the LCFA underestimates the production yield and loses completely the polarized contribution in circularly polarized laser pulses.

\ST{As an outlook, one possible idea to measure the polarization property of the produced particles is to collide the produced high-energy particle beam with an intense laser pulse in the upcoming laser-particle experiments~\cite{Abramowicz:2021zja,slacref1} by observing the spin-dependent nonlinear Compton scattering probability~\cite{seipt18,SeiptPRA052805} and the spin-dependent azimuthal distribution of the scattered photons~\cite{PRApplied014047}.}

\section{Acknowledgments}
The author thanks B. King for helpful discussions and comments on the manuscript.
The author is supported by the Natural Science Foundation of China, Grants No.~12104428. 
The work was carried out at National Supercomputer Center in Tianjin, and the calculations were performed on TianHe-1(A).
\appendix

\section{Spin basis}\label{Definition_of_Bispinors}
One can define a covariant spin basis vector: lightfront helicity,
\begin{align}
S^{\mu}_{p,z}&=\frac{p^{\mu}}{m}-\frac{m}{k\cdot p}k^{\mu}\,.
\label{Eq_lightfront_helicity}
\end{align}
in the lab reference, corresponding to the spin quantization direction
\begin{align}
\bm{n}=\frac{p^{+}+m}{p^{+}(p^{0}+m)}(p^{1},p^{2},p^{3}+m\frac{p^{0}+m}{p^{+}+m})\,,
\end{align}
which is antiparallel to the laser propagating direction in the particle rest frame, and where $p^{+}=p^{0}+p^{3}=k\cdot p/k^{0}$
The meaning of the subscript `$z$' in~(\ref{Eq_lightfront_helicity}) becomes clear when consider small-angle collisions ($\theta\ll1$) with $p_{z}\gg p_{x},p_{y}$, which is typical in the upcoming laser-particle experiments~\cite{Abramowicz:2021zja,slacref1}, the spin quantization, in the particle rest frame, points approximately to the $z$-direction as $\bm{n}\approx \left(0,0,1\right)$.

The spin eigenfunctions along the quantization direction $\bm{n}$ can be calculated via
\[\hat{\bm{\sigma}}\bm{n}w_{\bm{n},\sigma}=\sigma w_{\bm{n},\sigma}\,,\]
where $\hat{\bm{\sigma}}=(\hat{\sigma}_{1},\hat{\sigma}_{2},\hat{\sigma}_{3})$ are the Pauli matrices, and can be expressed explicitly as
\begin{subequations}
\begin{align}
w_{\bm{n},+1}&=\frac{1}{\sqrt{2p^{+}(p^{0}+m)}}\begin{bmatrix}
   p^{+}+m\\
   p^{1}+ip^{2}
\end{bmatrix}\,,\\
w_{\bm{n},-1}&=\frac{1}{\sqrt{2p^{+}(p^{0}+m)}}\begin{bmatrix}
   ip^{2}-p^{1}\\
   p^{+}+m
\end{bmatrix}\,.
\end{align}
\end{subequations}
The matrix elements of the Pauli matrices between the states $w_{\bm{n},\sigma}$ are given by:
\begin{align}
w^{\dagger}_{\bm{n},\sigma'}\hat{\bm{\sigma}} w_{\bm{n},\sigma}=\begin{bmatrix}
    \bm{n}     & \bm{e}_1-i\bm{e}_2 \\
    \bm{e}_1+i\bm{e}_2&  -\bm{n}
\end{bmatrix}_{\sigma'\sigma}=\bm{\Sigma}_{\sigma'\sigma}\,,
\end{align}
where $\bm{\Sigma}_{\sigma,\sigma}=\sigma\bm{n}$, $\bm{\Sigma}_{-\sigma,\sigma}=\bm{e}_1+ i\sigma\bm{e}_2$, and
\begin{align}
\bm{e}_1=&(1,0,0)-\frac{p^{1}}{p^{+}(p^{0}+m)}(p^{1},p^{2},p^{+}+m)\,,\nonumber\\
\bm{e}_2=&(0,1,0)-\frac{p^{2}}{p^{+}(p^{0}+m)}(p^{1},p^{2},p^{+}+m)\,,\nonumber
\end{align}
fulfilling that $\bm{n}\cdot \bm{e}_{1,2}=0$, $\bm{e}_{1}\cdot\bm{e}_{2}=0$, and $\bm{e}_{1}\times\bm{e}_{2}=\bm{n}$. One can also have spin density matrix in the particle rest frame as
\begin{align}
w_{\bm{n},\sigma} w^{\dagger}_{\bm{n},\sigma'} =  \frac{1}{2}\left(\mathbb{1}_{2\times 2} \delta_{\sigma\sigma'} + \hat{\bm{\sigma}}\bm{\Sigma}_{\sigma'\sigma}\right)\,.
\end{align}

In terms of the Chiral (Weyl) representation~\cite{nagashima2011elementary,schwartz2014quantum,peskin2018introduction}:
\begin{align}
\gamma^{0}&=\begin{bmatrix}
    0 & \mathbb{1}_{2\times2}  \\
    \mathbb{1}_{2\times2}  & 0
\end{bmatrix}\,,~
\gamma^{1}=\begin{bmatrix}
    0 & \hat{\sigma}_1 \\
   -\hat{\sigma}_1 & 0
\end{bmatrix}\,,~\nonumber\\
\gamma^{2}&=\begin{bmatrix}
    0 & \hat{\sigma}_2 \\
   -\hat{\sigma}_2 & 0
\end{bmatrix}\,,~
\gamma^{3}=\begin{bmatrix}
    0 & \hat{\sigma}_3 \\
   -\hat{\sigma}_3 & 0
\end{bmatrix}\,,\nonumber
\end{align}
the bispinors of electron ($u_{p,\sigma}$) and positron ($v_{p,\sigma}$), in the lab reference where the particles have momentum $p$, can be acquired as~\cite{nagashima2011elementary}:
\begin{subequations}
\begin{align}
u_{p,\sigma}&=\sqrt{\frac{p^{0}+m}{4m_0}}\begin{bmatrix}
w_{\bm{n},\sigma}-\frac{\hat{\bm{\sigma}}\bm{p}}{p^{0}+m}w_{\bm{n},\sigma} \\
w_{\bm{n},\sigma}+\frac{\hat{\bm{\sigma}}\bm{p}}{p^{0}+m}w_{\bm{n},\sigma}
\end{bmatrix}\,,\\
v_{p,\sigma}&=\sqrt{\frac{p^{0}+m}{4m}}\begin{bmatrix}
 \tilde{w}_{\bm{n},\sigma}-\frac{\hat{\bm{\sigma}}\bm{p}}{p^{0}+m}\tilde{w}_{\bm{n},\sigma} \\
-\tilde{w}_{\bm{n},\sigma}-\frac{\hat{\bm{\sigma}}\bm{p}}{p^{0}+m}\tilde{w}_{\bm{n},\sigma}
\end{bmatrix}\,,
\end{align}
\end{subequations}
where $\tilde{w}_{\bm{n},\sigma}=i\hat{\sigma}_2w^{*}_{\bm{n},\sigma}$. $v_{p,\sigma}=\hat{C}\gamma^{0} u^{*}_{p,\sigma}$ satisfies clearly the charge conjugation symmetry~\cite{bulbul2000relativistic}, and $\hat{C}=i\gamma^{2}\gamma^{0}$. Both the spinors fulfill the free Dirac equation as
\[
(\slashed{p}-m)u_{p,\sigma}=0\,,~~
(\slashed{p}+m)v_{p,\sigma}=0\,,
\]
and the relations:
\[
\bar{u}_{p,\sigma'}u_{p,\sigma}=\delta_{\sigma,\sigma'}\,,~
\bar{v}_{p,\sigma'}v_{p,\sigma}=-\delta_{\sigma,\sigma'}\,,~
\bar{u}_{p,\sigma'}v_{p,\sigma}=0\nonumber\,.
\]
The covariant spin density matrix can be given by~\cite{PRD016005}
\begin{subequations}
\begin{align}
\bar{u}_{p,\sigma}\gamma^{\mu} \gamma^{5}u_{p,\sigma'}&=S^{\mu}_{p,\sigma\sigma'}\,,\\
\bar{v}_{p,\sigma}\gamma^{\mu}\gamma^{5} v_{p,\sigma'}&=-S^{\mu}_{p,\sigma'\sigma}\,,
\end{align}
\end{subequations}
where $\gamma^{5}=i\gamma^{0}\gamma^{1}\gamma^{2}\gamma^{3}$, and $S^{\mu}_{p,\sigma\sigma'}$ are the covariant spin basis vectors in the lab reference with the explicit expression
\begin{subequations}
\begin{align}
S_{p,+1+1} &=-S_{p,-1-1}=S_{p,z}\,,\\
S_{p,-\sigma\sigma} &= S_{p,x} + i\sigma S_{p,y}\,,
\end{align}
\label{Eq_4D_spin_basis}
\end{subequations}
where $k\cdot S_{p,\sigma'\sigma}=\delta_{\sigma,\sigma'}\sigma k\cdot p/m$, and
\begin{subequations}
\begin{align}
S_{p,x} =& \left(\frac{p^{1}}{p^{+}},1,0,-\frac{p^{1}}{p^{+}}\right)=\epsilon^{\mu}_{x}-\frac{p\cdot \epsilon_{x}}{k\cdot p}k^{\mu}\,,\\
S_{p,y} =& \left(\frac{p^{2}}{p^{+}},0,1,-\frac{p^{2}}{p^{+}}\right)=\epsilon^{\mu}_{y}-\frac{p\cdot \epsilon_{y}}{k\cdot p}k^{\mu}\,.
\end{align}
\end{subequations}
where $\epsilon^{\mu}_{x}=(0,1,0,0)$ and $\epsilon^{\mu}_{y}=(0,0,1,0)$. One can also calculate the outer products of the bispinors~\cite{DIEHL200341,PRD016005} as
\begin{subequations}
\begin{align}
u_{p,\sigma}\bar{u}_{p,\sigma'}&=\frac{\slashed{p}+m}{4m} \left(\delta_{\sigma\sigma'}+\gamma^{5}\slashed{S}_{p,\sigma'\sigma}\right)\,,\\
v_{p,\sigma}\bar{v}_{p,\sigma'}&=\frac{\slashed{p}-m}{4m} \left(\delta_{\sigma\sigma'}+\gamma^{5}\slashed{S}_{p,\sigma\sigma'} \right)\,,
\end{align}
\end{subequations}
and the expectation value of the Pauli-Lubanski pseudovector,
\[W^{\mu}=-\frac{1}{4}\epsilon^{\mu\nu\alpha\beta}\sigma_{\nu\alpha}p_{\beta}\,,\]
where $\sigma_{\nu\alpha}=i[\gamma_{\nu},\gamma_{\alpha}]/2=i\gamma_{\nu}\gamma_{\alpha}$, as
\begin{align}
\bar{u}_{p,\sigma}W^{\mu}u_{p,\sigma'}=&\frac{m}{2}S^{\mu}_{p,\sigma,\sigma'}\,,\\
\bar{v}_{p,\sigma}W^{\mu}v_{p,\sigma'}=&\frac{m}{2}S^{\mu}_{p,\sigma',\sigma}\,.
\end{align}
The explicit expression of the bispinors are given by:
\begin{align}
 u_{p,+1}&=\frac{1}{\sqrt{2mp^{+}}}\begin{bmatrix}
   m\\
   0\\
   p^{+}\\
   p^{1}+ip^{2}
\end{bmatrix}\,,\nonumber\\
u_{p,-1}&=\frac{1}{\sqrt{2m p^{+}}}\begin{bmatrix}
   ip^{2}-p^{1}\\
   p^{+}\\
   0\\
   m
\end{bmatrix}\,,\nonumber\\
v_{p,+1}&=\frac{1}{\sqrt{2mp^{+}}}\begin{bmatrix}
    p^{1}-ip^{2} \\
   -p^{+} \\
    0 \\
    m
\end{bmatrix}\,,\nonumber\\
v_{p,-1}&=\frac{1}{\sqrt{2 m p^{+}}}\begin{bmatrix}
 m \\
 0 \\
 -p^{+}\\
 -(p^{1}+ip^{2})
\end{bmatrix}\,.\nonumber
\end{align}

\subsection{Polarization degree}
The electron spin density matrix with the arbitrary polarization $\bm{\Xi}_{\LCm}$ can be expressed as
\begin{align}
\rho^{\LCm}=\frac{1}{2}\left(\mathbb{1}_{2\times2}+\bm{\sigma}\bm{\Xi}_{\LCm}\right)
\label{Eq_density_matrix_electron}
\end{align}
where $|\bm{\Xi}_{-}|=1$ corresponds to a pure state polarized in the direction $\bm{\Xi}_{-}$, and $|\bm{\Xi}_{-}|<1$ corresponds to a mixed state polarized in the direction $\bm{\Xi}_{-}/|\bm{\Xi}_{-}|$. Analogously, the positron spin density matrix with the arbitrary polarization $\bm{\Xi}_{\LCp}$ can be expressed as
\begin{align}
\rho^{\LCp}=\frac{1}{2}\left(\mathbb{1}_{2\times2}+\bm{\sigma}\bm{\Xi}_{\LCp}\right)
\label{Eq_density_matrix_positron}
\end{align}
where $|\bm{\Xi}_{\LCp}|\leq1$. 

With the spin polarization density matrices~(\ref{Eq_density_matrix_electron}) and (\ref{Eq_density_matrix_positron}), one can simply calculate the covariant spin vector of the electron and positron; For the electron (pure) state $u_{p}=c_{\LCp}u_{p,\LCp}+c_{\LCm}u_{p,\LCm}$ as an example where $|c_{\LCp}|^{2}+|c_{\LCm}|^{2}=1$, one can obtain its covariant spin vector $S^{\mu}_{p}$ by calculating the expectation value of the Pauli-Lubanski operator as
\begin{align}
S^{\mu}_{p}=&\frac{2}{m}\bar{u}_{q}W^{\mu} u_{q}=\trm{Tr}\left[
\rho^{\LCm}
\left(
  \begin{array}{cc}
    S^{\mu}_{p,\LCp,\LCp} & S^{\mu}_{p,\LCp,\LCm} \\
    S^{\mu}_{p,\LCm,\LCp} & S^{\mu}_{p,\LCm,\LCm} \\
  \end{array}
\right)
\right]\nonumber\\
=&\Xi_{\LCm,z} S^{\mu}_{p,z}
+ \Xi_{\LCm,x} S^{\mu}_{p,x}
+ \Xi_{\LCm,y} S^{\mu}_{p,y}\,,
\end{align}
where the spin polarization components $\Xi_{\LCm,x,y,z}$ can be obtained directly from the spin coefficients $\rho^{\LCm}_{\sigma,\sigma'} = c_{\sigma}c^{*}_{\sigma'}$. At the low-energy limit $p^{\mu}\approx(m,0,0,0)$, the electron spin vector would go back to the unrelativistic case as \mbox{$S^{\mu}_{p}=(0,~\Xi_{\LCm,x},~\Xi_{\LCm,y},~\Xi_{\LCm,z})$}.
The analogous calculation can also be done for the spin vector of positron following the hole theory~\cite{bulbul2000relativistic};
the positive-energy positron should be regarded as a negative-energy electron in the spin state $v_{q}=\hat{C}\gamma^{0} u^{*}_{q}=c^{*}_{\LCp}v_{q,\LCp} + c^{*}_{\LCm}v_{q,\LCm}$, its covariant spin vector $S^{\mu}_{q}$ can be obtained by calculating the expectation value of the Pauli-Lubanski operator as
\begin{align}
S^{\mu}_{q}=&\frac{2}{m}\bar{v}_{q}W^{\mu}v_{q}=\trm{Tr}\left[
\rho^{*}_{\LCp}
\left(
  \begin{array}{cc}
    S^{\mu}_{q,\LCp,\LCp} & S^{\mu}_{q,\LCm,\LCp} \\
    S^{\mu}_{q,\LCp,\LCm} & S^{\mu}_{q,\LCm,\LCm} \\
  \end{array}
\right)
\right]\nonumber\\
=&\Xi_{\LCp,z} S^{\mu}_{q,z}
+ \Xi_{\LCp,x} S^{\mu}_{q,x}
+ \Xi_{\LCp,y} S^{\mu}_{q,y}\,,
\end{align}
where the spin polarization components $\Xi_{\LCp, x,y,z}$ can be obtained from the factors $\rho^{\LCp}_{\sigma,\sigma'} = c_{\sigma}c^{*}_{\sigma'}$. 

\section{Polarization basis and Stokes parameters}\label{photon_pol_density}
One can define the lightfront circular polarization basis:
\begin{subequations}
\begin{align}
|+\rangle:~\varepsilon^{\mu}_{+}=&\epsilon^{\mu}_{+}-\frac{\ell\cdot \epsilon_{+}}{k\cdot \ell}k^{\mu}\,,\\
|-\rangle:~\varepsilon^{\mu}_{-}=&\epsilon^{\mu}_{-}-\frac{\ell\cdot \epsilon_{-}}{k\cdot \ell}k^{\mu}\,,
\end{align}
\end{subequations}
in terms of the photon momentum ($\ell^{\mu}$) and laser wavevector [$k^{\mu}=\omega(1,0,0,-1)$], where \mbox{$\epsilon_{\pm}=(\epsilon_{x}\pm i\epsilon_{y})/\sqrt{2}$} and $\epsilon_{x}=(0,1,0,0)$, $\epsilon_{y}=(0,0,1,0)$. One can clearly have $\ell\cdot \varepsilon_{\pm}=0$ and $k\cdot \varepsilon_{\pm}=0$.

Because of the consideration for small-angle collisions ($\theta\ll1$) between high-energy seed photons and laser pulses: \mbox{$(\ell\cdot \epsilon_{\pm})/(k\cdot \ell)k^{\mu}\sim k^{\mu}\sin\theta /2\omega\sim \theta/2$}, one can approximate \mbox{$\varepsilon^{\mu}_{\pm}\approx \epsilon^{\mu}_{\pm}$}. Therefore, the basis $|\pm\rangle$ can be understood as the left/right-hand polarization states with the helicity value $+1$ and $-1$ respectively.

Any other polarization can be given as the linear superposition of the circular basis, \emph{e.g.}
\begin{subequations}
\begin{align}
|x\rangle&=\frac{ 1}{\sqrt{2}} \left(|+\rangle +|-\rangle\right):~\varepsilon^{\mu}_{x}=\epsilon^{\mu}_{x}-\frac{\ell\cdot \epsilon_{x}}{k\cdot \ell}k^{\mu}\,, \\
|y\rangle&=\frac{-i}{\sqrt{2}} \left(|+\rangle -|-\rangle\right):~\varepsilon^{\mu}_{y}=\epsilon^{\mu}_{y}-\frac{\ell\cdot \epsilon_{y}}{k\cdot \ell}k^{\mu}\,,
\end{align}
\end{subequations}
denote respectively the states polarized in the $x$- and $y$-directions, and
\begin{subequations}
\begin{align}
| 45^{\circ}\rangle&=\frac{ 1}{\sqrt{2}} \left(|x\rangle + |y\rangle\right)\,,\\
|135^{\circ}\rangle&=\frac{-1}{\sqrt{2}} \left(|x\rangle - |y\rangle\right)\,.
\end{align}
\end{subequations}
are the linear polarization states in the $x$-$y$ plane pointing along the azimuthal angle $\psi=45^{\circ}$ and $135^{\circ}$: \mbox{$|\psi\rangle=\cos\psi |x\rangle + \sin\psi|y\rangle$}.

A state with arbitrary polarization can be expressed via the density matrix theory in terms of the helicity states $|\pm\rangle$ as
\begin{align}
\rho_{\gamma}=\begin{bmatrix}
  \rho_{\gamma,++} & \rho_{\gamma,+-} \\
  \rho_{\gamma,-+} & \rho_{\gamma,--}
\end{bmatrix}
\end{align}
with the elements $ \rho_{\gamma,\sigma\sigma'}=\langle\sigma|\rho_{\gamma}|\sigma'\rangle$.
The photon number (or intensity) can be given as $I=\textrm{Tr}(\rho_{\gamma})=\rho_{\gamma,++}+\rho_{\gamma,--}$. To discuss the polarization degree, one can use the Stokes parameters ($\Gamma_{1}$, $\Gamma_{2}$, $\Gamma_{3}$):
\begin{enumerate}
 \item[$\Gamma_{1}$:] the degree of linear polarization indicating the preponderance of the polarization in $|x\rangle$ state over that in $|y\rangle$ state,
    \begin{align}
    \Gamma_{1} =&\frac{\langle x|\rho_{\gamma}|x\rangle-\langle y|\rho_{\gamma}|y\rangle}{I}=\frac{\rho_{\gamma,-+}+\rho_{\gamma,+-}}{I}\,,
    \end{align}
 \item[$\Gamma_{2}$:] the degree of linear polarization indicating the preponderance of the polarization in $|45^{\circ}\rangle$ state over that in $|135^{\circ}\rangle$ state,
    \begin{align}
    \Gamma_{2} =&\frac{\langle45^{\circ}|\rho_{\gamma}|45^{\circ}\rangle-\langle135^{\circ}|\rho_{\gamma}|135^{\circ}\rangle}{I}\nonumber\\
               =&i\frac{\rho_{\gamma,+-}-\rho_{\gamma,-+}}{I}\,,
    \end{align}
 \item[$\Gamma_{3}$:] the degree of circular polarization indicating the preponderance of the polarization in $|+\rangle$ state over that in $|-\rangle$ state,
    \begin{align}
    \Gamma_{3} =&\frac{\langle+|\rho_{\gamma}|+\rangle-\langle-|\rho_{\gamma}|-\rangle}{I}=\frac{\rho_{\gamma,++}-\rho_{\gamma,--}}{I}\,.
    \end{align}
\end{enumerate}
One can then rewrite the photon density matrix in terms of the Stokes parameters,
\begin{align}
\rho_{\gamma}=\begin{bmatrix}
  \rho_{\gamma,++} & \rho_{\gamma,+-} \\
  \rho_{\gamma,-+} & \rho_{\gamma,--}
\end{bmatrix}=\frac{I}{2}\begin{bmatrix}
  1+\Gamma_{3} & \Gamma_{1}-i\Gamma_{2} \\
  \Gamma_{1}+i\Gamma_{2} & 1-\Gamma_{3}
\end{bmatrix}
\end{align}

\section{Spin polarization of the electron}\label{Electron_result}
The spin polarization vector of the produced electron is given as
\begin{subequations}
\begin{align}
\Xi_{\LCm,x}&=\frac{\hat{\digamma}}{\trm{P}}\left(-\frac{\Delta_{y}}{2t} - \Gamma_{1}\frac{\Delta_{y}}{2 s} + \Gamma_{2}\frac{\Delta_{x}}{2 s} + \Gamma_{3}\frac{\Sigma_{x}}{2 t} \right)\,,\\
\Xi_{\LCm,y}&=\frac{\hat{\digamma}}{\trm{P}}\left( \frac{\Delta_{x}}{2t} - \Gamma_{1}\frac{\Delta_{x}}{2 s} - \Gamma_{2}\frac{\Delta_{y}}{2 s} + \Gamma_{3}\frac{\Sigma_{y}}{2 t}  \right)\,,\\
\Xi_{\LCm,z}&=\frac{\hat{\digamma}}{\trm{P}}\left(\Gamma_{3}\frac{2-t\bm{\Delta}^{2} g_s}{2t} + ig_s \bm{w}(\phi_{1})\times\bm{w}(\phi_{2})\right).
\end{align}
\label{Eq_electron_dmatrix}
\end{subequations}
After integrating over the transverse momenta, one can arrive at the lightfront momentum distribution of the electron spin polarization vector, as
\begin{subequations}
\begin{align}
\Xi_{\LCm,x}(t)&=\frac{\hat{T}}{\ud\trm{P}/\ud t} \left(\frac{\Gamma_{3} \mathfrak{b}_{x} - \Delta_{y}}{2t}
 -\frac{ \Gamma_{1}\Delta_{y} -\Gamma_{2}\Delta_{x}}{2 s}\right),\\
\Xi_{\LCm,y}(t)&=\frac{\hat{T}}{\ud\trm{P}/\ud t} \left(\frac{\Gamma_{3} \mathfrak{b}_{y} + \Delta_{x}}{2t}
 -\frac{\Gamma_{1}\Delta_{x} + \Gamma_{2}\Delta_{y}}{2 s}\right),\\
\Xi_{\LCm,z}(t)&=\frac{\hat{T}}{\ud\trm{P}/\ud t} \left[\Gamma_{3}\left(\frac{1}{t} - g_{s}\frac{\bm{\Delta}^{2}}{2 }\right) +i g_{s} \bm{\mathfrak{a}}(\phi_{1})\times \bm{\mathfrak{a}}(\phi_{2})\right],
\end{align}
\end{subequations}
where $\ud\trm{P}/\ud t$ is the energy spectrum of the electron, same as the one of the positron with the exchange $s\leftrightarrow t$.

The LMA result for the electron's spin polarization vector in the circularly polarized laser pulse is given as
\begin{subequations}
\begin{align}
\Xi_{\LCm,x}(t)&=0\,,\\
\Xi_{\LCm,y}(t)&=0\,,\\
\Xi_{\LCm,z}(t)&=\frac{\alpha/\eta }{\ud\trm{P}/\ud t}\int \ud\vphi \sum_{n=\lceil n_{\ast} \rceil} \left\{2c \xi(\vphi) g_{s} J_{n} J'_{n} \left( \frac{n\eta t s}{r_{n}}-r_{n}\right) \right.\nonumber\\
              +&\left.\Gamma_{3}\left[J^2_{n}/t - g_{s}\xi^2\left(n^2 J^2_{n}/\zeta^2_{n} + J'^2_{n} - J^2_n\right)\right]\right\},
\end{align}
\end{subequations}
and in the linearly polarized laser pulse, as
\begin{subequations}
\begin{align}
\Xi_{\LCm,x}(t)=&0\,,\\
\Xi_{\LCm,y}(t)=&0\,,\\
\Xi_{\LCm,z}(t)=&\frac{\Gamma_{3}}{\ud \trm{P}/\ud t}\frac{\alpha}{ \eta }  \int \ud\vphi \int^{\pi}_{-\pi}\frac{\ud \theta}{2\pi}  \sum_{n=\lceil n_{\ast} \rceil} \nonumber\\
                & \left[\Lambda^{2}_{0,n}/t - \xi^{2}(\vphi)g_{s}\left(\Lambda^{2}_{1,n}- \Lambda_{0,n} \Lambda_{2,n}\right)\right] \,.
\end{align}
\end{subequations}

The LCFA result for the electron's spin vector is also provided as
\begin{subequations}
\begin{align}
\Xi_{\LCm,x}(t)&=\frac{-\alpha/\eta }{\ud\trm{P}/\ud t} \int\frac{\ud\vphi}{\sqrt{\zeta}}\left(\frac{\Gamma_{1}\xi_{y} -\Gamma_{2}\xi_{x}}{s|\xi|} + \frac{\xi_{y}}{t |\xi|} \right)\textrm{Ai}(\zeta)\,,\\
\Xi_{\LCm,y}(t)&=\frac{-\alpha/\eta }{\ud\trm{P}/\ud t} \int \frac{\ud\vphi}{\sqrt{\zeta}}\left(\frac{\Gamma_{1}\xi_{x} +\Gamma_{2}\xi_{y}}{|\xi|s} - \frac{\xi_{x}}{t |\xi|} \right)\textrm{Ai}(\zeta)\,,\\
\Xi_{\LCm,z}(t)&=\frac{\alpha/\eta }{\ud\trm{P}/\ud t} \int \ud\vphi \left[ \frac{1}{t}\textrm{Ai}_1(\zeta) + \frac{2}{\zeta}\textrm{Ai}'(\zeta)g_{s} \right] \Gamma_{3}\,.
\end{align}
\end{subequations}

Compared to the corresponding results for the produced positron, one can clearly see the symmetry between the paired particles by exchanging their lightfront momenta and considering their charge with different sign.

\bibliographystyle{apsrev}

\end{document}